\documentclass[10pt,journal,compsoc]{IEEEtran}
\ifCLASSOPTIONcompsoc
  \usepackage[nocompress]{cite}
\else
  \usepackage{cite}
\fi

\ifCLASSINFOpdf
\else
\fi

\usepackage{url}
\usepackage{graphicx}
\usepackage{pgfplots}
    \pgfplotsset{compat=1.3}
\usepackage{tikz}
\usetikzlibrary{patterns}

\begin{document}
\title{Exploring Emotions in Multi-componential Space using Interactive VR Games}

\author{Rukshani~Somarathna,~\IEEEmembership{Member,~IEEE,}
        and~Gelareh~Mohammadi,~\IEEEmembership{Member,~IEEE}
\IEEEcompsocitemizethanks{\IEEEcompsocthanksitem Rukshani Somarathna is with the School of Computer Science and Engineering, University of New South Wales, Australia, e-mail: r.somarathna@student.unsw.edu.au.
\IEEEcompsocthanksitem Gelareh Mohammadi is with the School of Computer Science and Engineering, University of New South Wales, Australia, e-mail: g.mohammadi@unsw.edu.au.}
}


\IEEEtitleabstractindextext{%
\begin{abstract}
Emotion understanding is a complex process that involves multiple components. The ability to recognise emotions not only leads to new context awareness methods but also enhances system interaction's effectiveness by perceiving and expressing emotions. Despite the attention to discrete and dimensional models, neuroscientific evidence supports those emotions as being complex and multi-faceted. One framework that resonated well with such findings is the Component Process Model (CPM), a theory that considers the complexity of emotions with five interconnected components: appraisal, expression, motivation, physiology and feeling. However, the relationship between CPM and discrete emotions has not yet been fully explored. Therefore, to better understand emotions’ underlying processes, we operationalised a data-driven approach using interactive Virtual Reality (VR) games and collected multimodal measures (self-reports, physiological and facial signals) from 39 participants. We used Machine Learning (ML) methods to identify the unique contributions of each component to emotion differentiation. Our results showed the role of different components in emotion differentiation, with the model including all components demonstrating the most significant contribution. Moreover, we found that at least five dimensions are needed to represent the variation of emotions in our dataset. These findings also have implications for using VR environments in emotion research and highlight the role of physiological signals in emotion recognition within such environments.
\end{abstract}

}

\maketitle

\IEEEdisplaynontitleabstractindextext
\IEEEpeerreviewmaketitle

\IEEEraisesectionheading{\section{Introduction}\label{sec:introduction}}

\IEEEPARstart{U}{nderstanding} emotions and their formation is a fundamental aspect of human social interactions, and it has extensive implications in many fields, such as psychology, computer science, and human-computer interaction. Emotions are complex cultural and psychobiological states that are an important aspect of human experience and play a central role in our social interactions and decision-making \cite{RN20}. Therefore, emotion understanding enables individuals to comprehend and respond appropriately to events and regulate their emotions. Additionally, research on emotion understanding has the potential to inform the development of systems that possess context awareness and the ability to perceive and express emotions \cite{RN632}. Recent developments in the field have led to numerous innovations and advances in Artificial Intelligence (AI) systems such as virtual assistance, chatbots, Virtual Reality (VR) environments, cross-cultural models that can recognise and respond to emotions across different cultures, wearable devices, and smart applications that benefit from emotional awareness. Although discrete and dimensional models have received significant attention, neuroscientific evidence suggests that emotions are complex, multi-faceted in nature, and show multiple brain processes subserving emotional experience \cite{RN676, RN645}. Affective Computing (AC) research is mainly attributed to Discrete, Dimensional, and Appraisal \cite{RN182, RN79} models in data-driven analysis. Each of these models approaches emotions from different perspectives; however, only discrete, and dimensional models have gained widespread recognition in research on emotion formation \cite{RN22, RN473}. Discrete models suggest that emotions are distinct, separate entities that can be easily distinguished from one another \cite{RN23, RN22}. Dimensional models postulate that emotions can be differentiated along continuous dimensions such as valence and arousal \cite{RN73, RN182, RN22, RN28}. 

Appraisal models theorise that emotions are the responses of an individual's cognitive evaluation of an event \cite{RN295, RN471, RN607}. Appraisal models assume the cognitive processes, the role of individual differences, and contextual awareness. Nevertheless, discrete models may not adequately capture the complexity of emotion formation, as they treat emotions as distinct categories \cite{RN247, RN131} rather than considering the continuity of emotional experiences. Dimensional models address this to some extent by recognising the continuity, however mostly limited to basic dimensions \cite{RN182, RN22}. Further, both discrete and dimensional models tend to focus on the feeling component without explaining the interactive effects of sub-processes \cite{RN20}. While appraisal models acknowledge the role of cognitive processes in emotion formation, there is limited data-driven research \cite{RN22, RN67, RN471}. One possible reason may be the complexity of operationalising and assessing, as they rely on subjective evaluations of events. However, as the appraisal model recognises the complexity of emotions, encompassing a range of cognitive, physiological, and behavioural processes \cite{RN182, RN21}, it is crucial to consider a process-based model to understand the mechanisms behind emotion formation. 

The appraisal model has influenced a significant amount of research. The Component Process Model (CPM) \cite{RN295, RN21, RN20}, a variant of the appraisal model, has also significantly shaped this research. CPM assumes the process-based modelling between five main components: appraisal, motivation, expression, physiology, and feeling \cite{RN20}. Renewed interest in the CPM has led to data-driven analyses. However, these studies have often been limited in their use of active participation \cite{RN590, RN295}, objective measures of the physiology and expression components \cite{RN67}, and the induction of a broader range of emotions \cite{RN97, RN295}. Despite a few existing studies on CPM, there is a need for more research using enhanced data-driven approaches that involve active participation and objective measures to gain a better understanding of emotions.

In this study, we operationalise a VR-based setting to explore the relationship between CPM and discrete emotions by creating a more realistic emotional induction \cite{RN650}. In this research, our emphasis was on CPM due to the availability of a standard questionnaire to measure components and operationalise them. We induced a more comprehensive range of emotions using interactive VR games and collected multimodal measures: self-reports, physiological, and facial signals. Our goal is to better understand the role of each CPM component in capturing emotions using self-reports and physiological and facial signals. We also explore the potential of VR games to induce a range of emotions with different intensities. It's worth noting the main reason for employing VR games is that they provide an immersive and more ecological medium for eliciting emotions, which goes beyond the previous studies that used still images or video clips that result in a more passive experience of emotions. Therefore, our analyses in this manuscript are confined to understanding emotions rather than the impact on VR gameplay. Additionally, we examine latent dimensions that help understand distinct emotional features using CPM self-reports. Our findings inform the role of each CPM component in capturing emotions. These insights can be used for enhanced context awareness, system interaction, and system adaptations in domains including healthcare, VR environments, education, gaming, and interfaces. Moreover, it uses a data-driven approach to investigate emotion formation positing that emotions are subjective experiences.

This manuscript builds upon previous research in this area \cite{RN649, RN646, RN701} by presenting more comprehensive experiments and analyses with further data samples and generalised ML models using more rigorous evaluation methods (e.g. Leave-One-Subject-Out (LOSO) cross-validations). This paper addresses the following research questions: (i) How are CPM and discrete emotions related when emotions are induced through interactive VR games? (ii) What are the latent dimensions that underlie emotional experience? and (iii) What is the contribution of each CPM component in capturing discrete emotions? This study makes several significant contributions to the field of affective computing and emotion recognition, which can be summarised as (i) Exploring the componential process model in an interactive VR environment with a much broader scope than previous works, (ii) Providing new insights into the relationship between discrete emotions and CPM, and (iii) Exploring the importance of each component in differentiating discrete emotions.

The paper is organised as follows: Section 2 presents theoretical foundations, Section 3 covers our research methodology, Section 4 details results and implications, and Section 5 concludes with a summary.

\section{Background and Related Works}
The purpose of this section is first to provide a background on the topic of emotion understanding from CPM and then to review relevant research in the field.

\subsection{Component Process Model (CPM)}
\begin{figure}[!t]
\centering
\includegraphics[width=\linewidth]{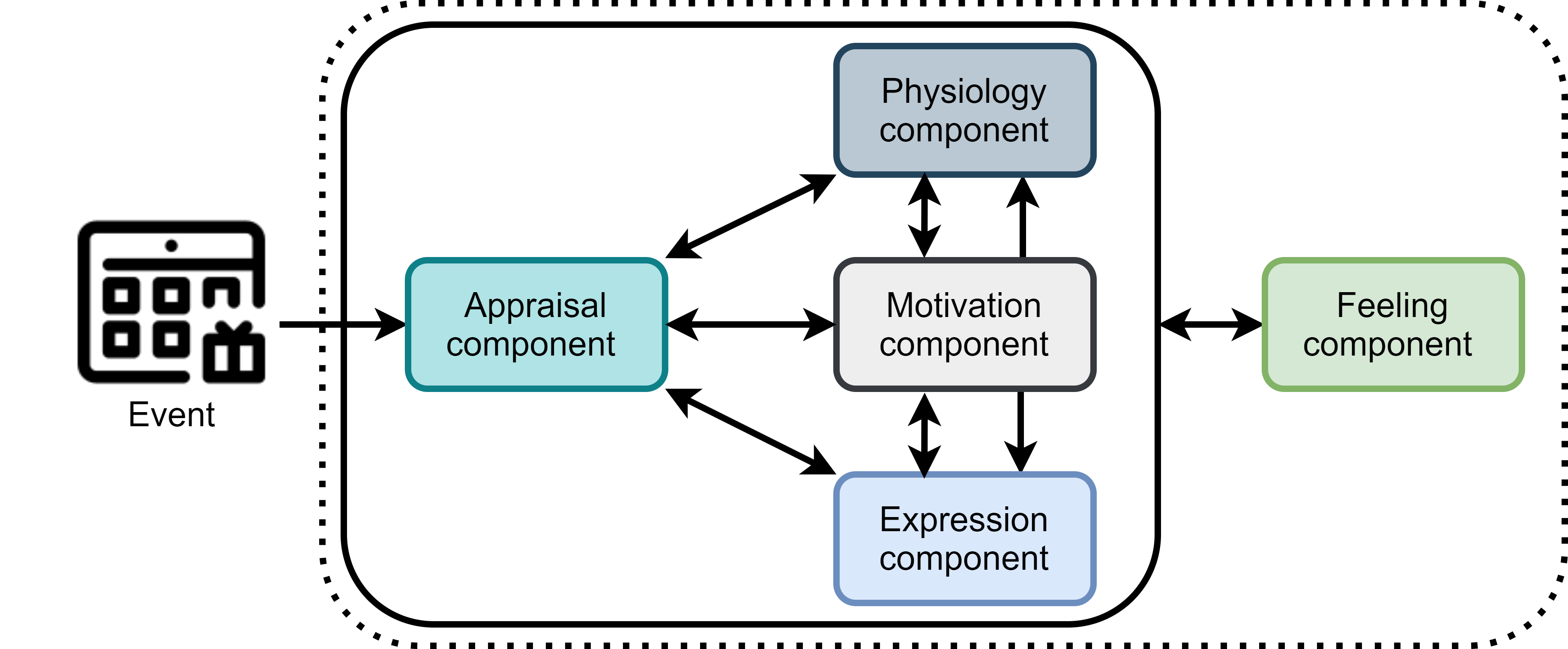}
\caption{Simple Component Process Model (CPM) based on \cite{RN21}.}
\label{fig:Figure1_SimpleCPM}
\end{figure}

The CPM is a theoretical framework composed of five main components: appraisal, motivation, physiology, expression, and feeling, which are used to understand and interpret emotional experiences \cite{RN21, RN20}. Refer to Figure \ref{fig:Figure1_SimpleCPM} for a visual representation of these components. CPM is based on the idea that emotions are complex and multi-faceted in nature, with various cognitive, physiological, and behavioural processes converging to create an emotional experience. 

The CPM's appraisal component consists of cognitive processes to evaluate an event. It has four objectives that evaluate an event as \emph{1) Relevance: ``Is this event relevant?'', 2) Implications: ``What will be the impact of this event?'', 3) Coping potential: ''What are the chances of me being able to overcome the potential consequences of this event?''} and \emph{4) Normative significance: ``Does this event align with my values and beliefs?''}. The outcome of these assessments activates mutually dependent processes on other components. The motivation component refers to the action tendencies, goals, and motivations triggered by an individual's appraisal of an event and drives their emotional reactions. Driven by appraisal and motivation outputs, the expression component initiates verbal or non-verbal behaviour to convey the emotional experience. Similarly, the physiology component begins appropriate changes in bodily functions to adjust to the situation. Finally, the feeling component represents the conscious and non-conscious subjective emotional experience. 

The suggestive features and functions of the CPM in emotion formation can be evaluated by the GRID instrument \cite{RN77} or by the shorter versions, CoreGRID and MiniGRID \cite{RN76}, which describe the meanings of emotion words. Unfolding the emotional experiences by CPM can be done by either selecting full or a combination of components. Accordingly, selecting proper GRID items that correspond to the chosen stimuli and selecting appropriate data-analytic procedures for operationalising the selected CPM components is critical \cite{RN77}.

\subsection{Related works}
In literature, CPM-based data-driven research aims to identify the structure of discrete emotions within the componential space using data. Further, the exploration of computational modelling in understanding emotions has been a recurring theme in diverse research endeavors \cite{marsella2010computational, RN607, yongsatianchot2021computational}. These approaches use collected data and computational models to provide bottom-up insights into discrete emotions from a CPM perspective. Its merit lies in minimising reliance on assumptions about the relationships between variables, enabling data to guide the interpretation of emotions from a CPM perspective.

Most of the data-driven research has been conducted using individual or a subset of CPM components \cite{RN69, RN479, RN578, RN97, RN269, RN34}. However, this narrow scope can also limit our comprehensive understanding of emotions, leading to less reliable conclusions being drawn by computational models. Therefore, considering a full CPM with five components can potentially improve the consistency of computational models and expand the current understanding of emotions \cite{RN21} and their representation within different components. For example, in a data-driven study by Mohammadi and Vuilleumier \cite{RN295}, the authors used the full CPM to investigate the relationship between 10 discrete emotions and their corresponding CoreGRID responses using films as stimuli. The researchers discovered a clear hierarchy between positive and negative emotions within the CPM space. They also found that six dimensions were necessary to capture the differences between emotions, and ML can be used to differentiate emotion features. Another data-driven study used VR to induce emotions actively and examined the full CPM \cite{RN67} but focused more on the appraisal component. Also, due to the limited selection of only 7 VR games, a broader range of emotions may not have been effectively induced. Additionally, their results were influenced by the novelty factor associated with experiencing VR, as not all participants had prior exposure to VR. Collectively, both of these data-driven studies have solely depended on measuring CoreGRID subjective annotations for interpreting the objectively measurable components (expression and physiology) of the CPM.

Physiological signals have been used in AC research \cite{RN54, RN28, RN83} to measure emotion responses, providing details about the non-conscious and conscious experience \cite{RN206, RN477}, intensity, valence, and temporal variations. Recent technological advancements have allowed it to collect physiological data non-invasively with minimal setup and configuration \cite{RN23, RN194, RN604}. Researchers have focused on using physiological signals to study different emotions, but relatively little attention has been given to using these signals in studies based on the full CPM. One such example is a data-driven study by Menétrey, et al. \cite{RN590}, which has been expanded to include heart activity, skin conductance, and respiration other than the CoreGRID annotations. Using the data collected inside a functional Magnetic Resonance Imaging (fMRI) machine, researchers confirmed ML's ability to differentiate emotions and assess synchronisation between the components during an emotional episode. The full CPM showed better when all components were considered rather than just one. However, this experiment setup may be less relaxing due to the fMRI setup and passive emotion induction due to the use of films \cite{RN650}. Moreover, some studies have used facial signals for measuring emotions.

Facial electromyography (fEMG) is a non-invasive technique that uses electrodes to measure facial muscle activations \cite{RN599, RN669}. Studies showed that fEMG can reflect changes in the dimensional model of emotion \cite{RN664, RN705, RN599, RN627}, discrete model of emotion \cite{RN494, RN666, RN610}, and facial expressions \cite{RN668, RN669, RN678}. Several studies have focused on individual components or processes of the CPM \cite{RN479, RN578}, but to our knowledge, none of these studies has examined the full CPM using fEMG as an objective measure of expression. 

Overall, the CPM is progressively getting more recognition in AC to understand the multi-faceted nature of emotions and underlying processes. The CPM can also be considered as a comprehensive model that takes context into account and goes beyond using only the expression. Current research has provided valuable insights into emotion formation's cognitive, physiological, and behavioural processes. However, further research is yet needed to provide deeper insights into underlying mechanisms, the role of physiology, and other components in a more immersive and naturalistic setting.  

Many of the previous studies in this field have utilised passive methods, such as films, for eliciting emotions. However, these methods have been shown to be less effective in eliciting emotions than active methods. While some studies have used active methods, such as VR, they have often restricted their stimulus selection to a narrower spectrum of emotions, relied solely on self-reporting, and have not fully investigated the objectively measurable components of emotions or the contributions of each component. Furthermore, most studies tend to overlook the subjective nature of emotions and pre-label emotional situations with fixed labels. Therefore, we propose a data-driven approach, postulating emotional experience as subjective, to understanding emotions from a componential perspective. We will employ active emotion elicitation techniques and collect multimodal objective and subjective measures to assess the contribution of each component in capturing discrete emotions.

\section{Methodology}
This section outlines the research design, material selection, and data collection procedures employed in this study.

\subsection{Material and Assessment}
Material selection is a significant step toward the reliability of research conclusions. Therefore, we used interactive VR games to create a more realistic and immersive emotion induction paradigm as opposed to passive paradigms \cite{RN703, RN650}. We selected 96 VR games from Steam\footnote{https://store.steampowered.com/} used in previous works \cite{RN30, RN67, RN605} and which were rated on game review websites. From that, we chose 27 VR games (refer to Supplementary Table S1) based on the game mechanics' simplicity, clarity, diversity, and the possibility to generate emotions within three minutes. For instance, the chosen games feature VR scenarios like shooting and zombie encounters mainly at evoking emotions such as fear, disgust, and hate. Additionally, VR experiences like slingshotting, slicing fruits, and slashing blocks are included to evoke feelings of joy, pleasure, and interest. To ensure the selection encompasses a wide range of emotions, games were labelled with a dominant emotion only for selection purposes using the Geneva Emotion Wheel (GEW)~\cite{RN190}; however, these labels were not used in the following analysis. We used all the 20 emotions from the GEW, which are defined as positive (interest, amusement, pride, joy, pleasure, contentment, admiration, love, relief, compassion) and negative (sadness, guilt, regret, shame, disappointment, fear, disgust, contempt, hate, anger). Given that VR games are primarily developed for entertainment purposes, it was difficult to identify games that could elicit negative emotions and some of the positive emotions within a three-minute time frame. Consequently, an uneven distribution of emotions was observed in the selected games, as indicated in Supplementary Table S1. 

We labelled seven games based on previous works \cite{RN30, RN67, RN605} and the rest of the 20 games based on researchers' gameplay experience and game review ratings and comments. While we termed each game with primary emotion, we did not expect all participants to experience the same emotions due to subjective variations \cite{RN97, RN706, RN83}. Therefore, we used the GEW and CoreGRID as surveys to collect participants' emotional experiences following each game. 

In the experimental setup, we organised a training session to train the participants to VR and to reduce the novelty effect of VR. For that, we used the SteamVR tutorial application, NVIDIA® VR Funhouse, Fantasynth One, Expedia Cenote VR, Luna, and Google Spotlight Stories: Age of Sail games. 

\subsection{Proposed system}
\begin{figure}
    \centering
    \includegraphics[trim=0.1cm 0.4cm 0.3cm 0cm, clip, width=\linewidth]{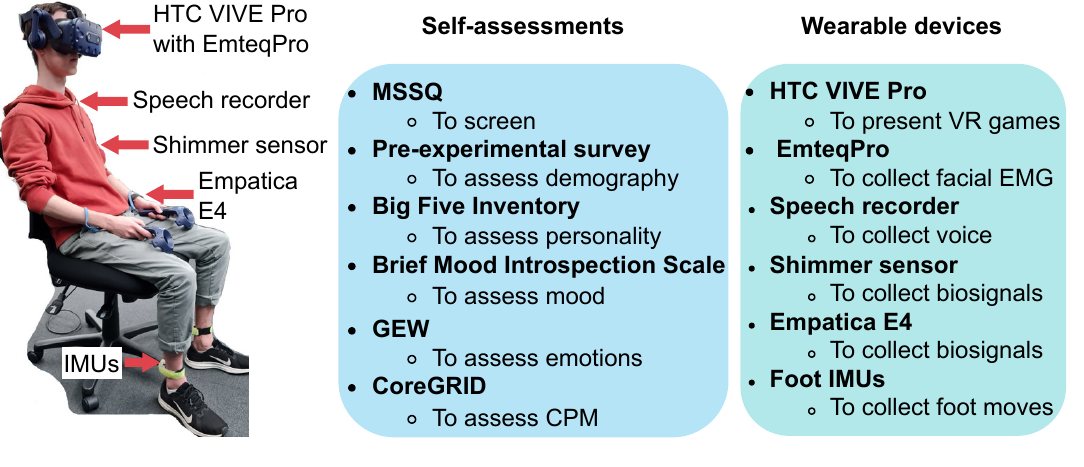}
    \caption{A participant playing VR games while wearing wearable devices, providing subjective assessments.}
    \label{fig:Figure2_Experiment}
\end{figure}

Figure \ref{fig:Figure2_Experiment} shows our experimental setup, where we used an HTC VIVE Pro headset with controllers for game presentation. We used emteqPRO \cite{RN660} wearable device to measure the facial EMG activations and Empatica E4 wristband to record Heart Rate (HR), Electrodermal Activity (EDA), Blood Volume Pulse (BVP), Inter Beat Intervals (IBI), skin temperature, and acceleration. We also used Shimmer sensors to measure respiration and Electrocardiography (ECG) and an off-the-shelf Inertial Measurement Unit (IMU) \cite{RN651} to track body movements. However, for the purpose of the present study, our focus was limited to the analysis of facial EMG, BVP, skin temperature, EDA, and respiration.

\subsection{Procedure}
The research was approved by the University of New South Wales, Human Research Ethics Committee (HC200809). The dataset was collected from 39 participants (18 females, 21 males, mean age = 25 \& SD = 5.42 years) who were first screened by the Motion Sickness Susceptibility Questionnaire (MSSQ) \cite{RN302} for motion sickness. The inclusion/exclusion criteria to recruit participants were an age range of 18-40 years, without any type of prior psychological or prior neurological disorders, fluent in the English language, not wearing glasses or with corrected refractive error, and without prior vertigo, hearing, or vestibular problems. Each participant went through a one-hour training session and three 100-minute actual data collection sessions as part of our data collection process. Each participant received \$150 as reimbursement for their participation. The training session began with an explanation of the experiment setup and instructions for the participants to report their own experiences rather than what they thought should be felt while playing the games in actual data collection sessions. 

We developed an app using PsychoPy \cite{RN702} to present the games, administer the questionnaires, synchronise each event, and save our data. To ensure a balanced distribution of emotional experiences across the three data collection sessions, we randomised the sequence of pre-labelled games. We minimised repeating any similar experiences within a single session. Additionally, we randomised the presentation of each survey item to reduce any effect of the order. We collected each participant's demography (pre-experimental survey), personality (Big Five Inventory survey), and mood (Brief Mood Introspection Scale survey) through surveys. 

For the calibration of wearable devices, we first collected neutral expressions. For the calibration of emteqPRO, we further collected several maximum smiles, frowns, and eyebrow raises expressions. To instruct the participants in the calibration phase, we developed 3D scenes using Blender \cite{RN699}, which were deployed to VR environments using HARFANG®3D \cite{RN698}. Then each participant played a VR game for three minutes and completed GEW and CoreGRID (refer to Supplementary Tables S2 and S3) surveys subsequently, where they marked emotional experience on a 5-point Likert scale (\emph{1-Not at all, 5-Strongly}). This process was repeated until each participant had completed all 27 games in three data collection sessions. We collected 1053 observations (27 games $\times$ 39 participants) from all participants but only considered 1041 observations in our analysis. The rest of the observations were removed due the technical concerns such as network errors, game updates, and headset disconnections. Lastly, we post-processed our collected facial EMG data using the SuperVision \cite{RN660} application to get the frequency and intensity of facial expressions such as smile, frown, eyebrow raises and neutral, arousal, and valence insights. 

\section{Results and Discussion}
The following subsections describe our results and implications. The first section, 4.1, discusses the different discrete emotions experienced in VR, while section 4.2 explores the componential features of these emotions. In section 4.3, the focus is on finding the emotional dimensions, while section 4.4 explains the role of facial EMG in the emotional experience. Finally, section 4.5 discusses the use of ML in modelling emotions and investigates the role of each component. Together, these sections provide a comprehensive understanding of the various facets of emotional experience in VR.

\subsection{Discrete Emotions in VR}
\begin{figure}
    \centering
    \includegraphics[trim=0.1cm 0.4cm 6.5cm 6.5cm, clip, width=1\linewidth]{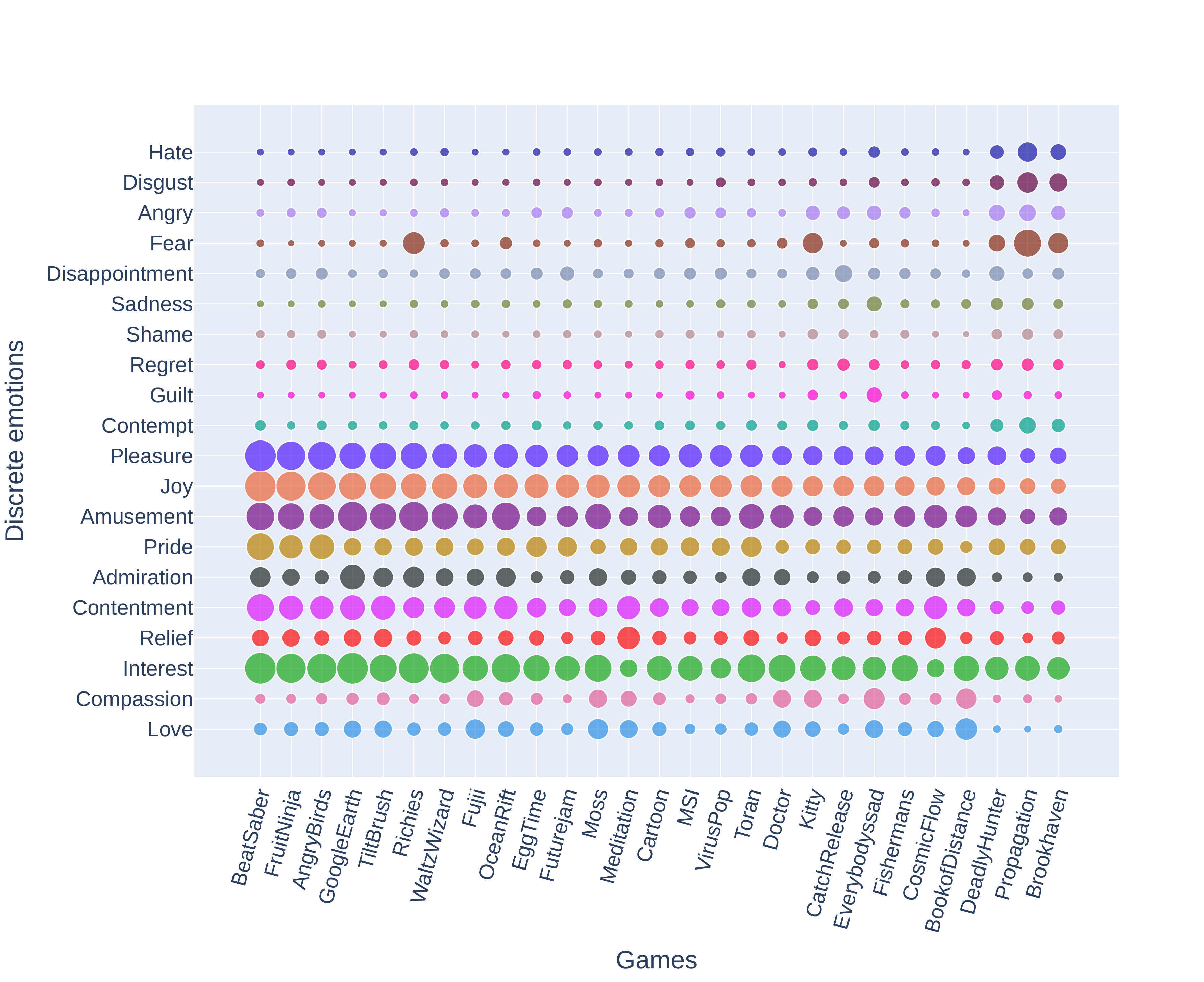}
    \caption{Participant's average intensity ratings for discrete emotions by VR games. The size of the bubble is proportional to the average emotional intensity. For better representation of the difference across games, the games were sorted based on the average rating of Joy emotion.}
    \label{fig:Figure3_BubbleEmoGamesSortedByJoy}
\end{figure}

Given the novelty of VR to emotional studies, the efficacy of VR games in inducing a wider range of emotions should be validated to answer our research questions. Therefore, in Figure \ref{fig:Figure3_BubbleEmoGamesSortedByJoy}, we illustrated the 39 participant's average intensity scores of each discrete emotion (defined by the 20 GEW terms) when playing the selected VR games. This plot may be important to researchers interested in selecting VR games that evoke specific emotions. It can provide insight into which VR games are most effective at eliciting specific emotions when designing future studies and aid in identifying which emotions are most elicited by VR games in general, which can inform the development of VR games that are more emotionally engaging. 

It is apparent from this figure that all the selected emotions were experienced to a certain intensity by participants. Regardless of the negative valence of some games, positive emotions such as interest, amusement, pride, joy, pleasure, contentment, and admiration are highly rated. This is supported by the high immersion and pleasantness in VR games that increase interaction and engagement. Additionally, love, relief, compassion, disappointment, fear, and anger have been triggered by some games. However, the intensity of sadness, guilt, regret, shame, disgust, contempt, and hate is relatively low. As VR games are generally targeted for entertainment, it was not easy to find games that elicit such complex emotions. Our results align with previous surveys \cite{RN650} and studies that found sadness, guilt, and contempt \cite{RN67} to be rated as low-intensity emotions in general in laboratory experiments. This could be due to the intricate nature of some of these emotions, which are often referred to as social or self-conscious emotions (shame, guilt, and contempt \cite{RN79, RN248, RN249}). Furthermore, we observed a discrepancy between our initial categorisation (Supplementary Table S1) of emotions, which was based on previous works, the experience of the researchers, and ratings and comments from game review sites. The outcomes of this analysis suggest that individuals exhibit varied emotions in similar situations. Consequently, the ratings for the selected games and their content reveal the presence and complex nature of appraisal objectives to react to an event adaptively. Moreover, the figure highlights the presence of complex emotional structures and mixed feelings in each game, which makes the current emotion recognition models that usually label an event with only one emotion less effective. The study's results indicate that our VR games are effective means of eliciting a range of emotions (though it is usually mixed emotions), which can be used to address our research questions.

\subsection{Componential Emotion Features in VR}
\begin{figure}
    \centering
    \includegraphics[width=1\linewidth]{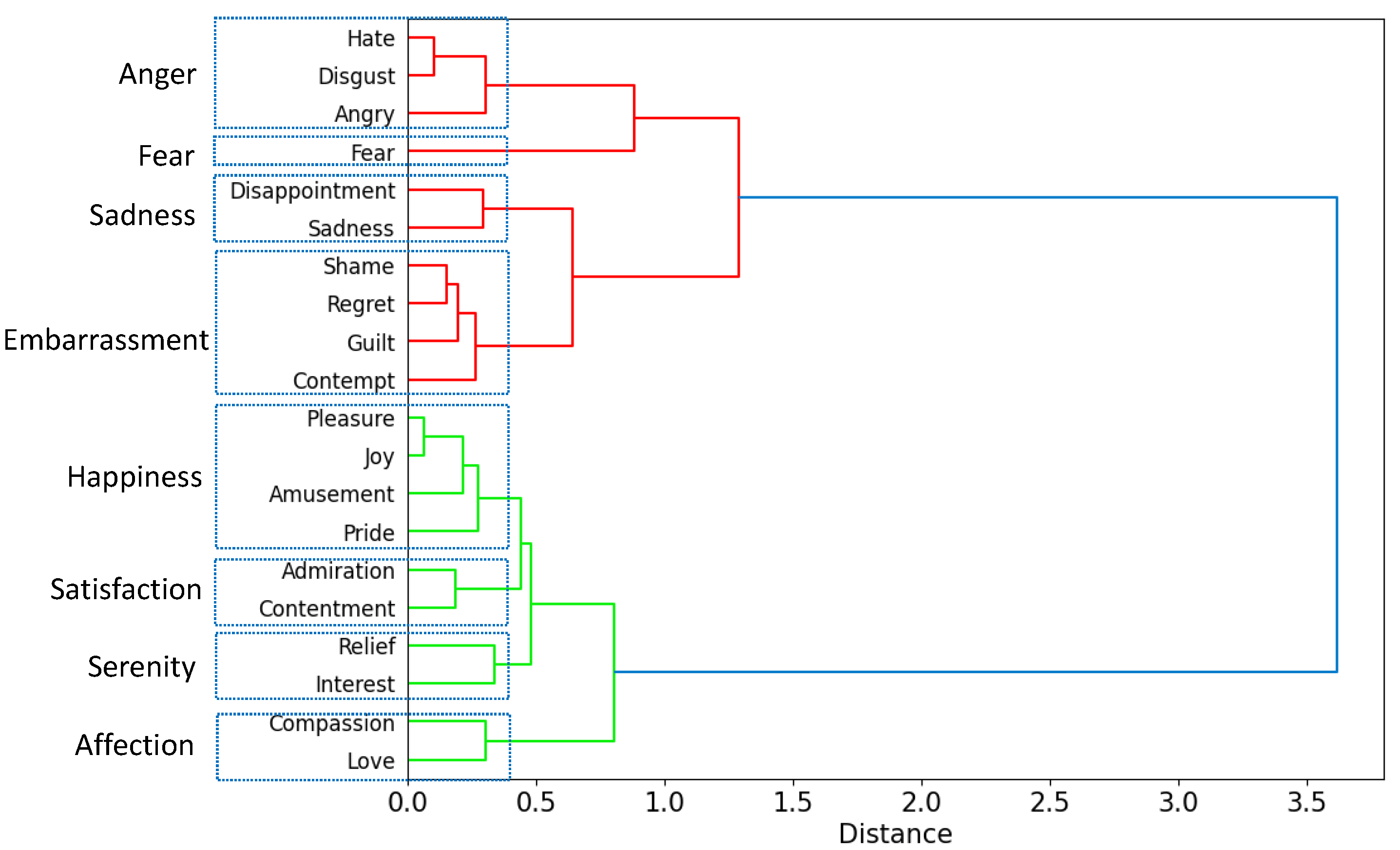}
    \caption{Results of the componential hierarchical clustering of discrete emotion terms. Two distinct clusters of negative (red cluster) and positive (green cluster) emotions with meaningful sub-clusters can be observed.}
    \label{fig:Figure4_CPMClusters}
\end{figure}

We performed a cluster analysis on the CPM profile of different emotion categories to unfold the hierarchical organisation of each emotion in the CPM space. Here, our analyses were based on the self-assessment of 51 CoreGRID and 20 GEW items, with the CoreGRID items serving as representatives within the CPM. The formula for the weighted average CPM profile (CP\textsubscript{j}) for emotion term \emph{j} is given as follows:

\[ {CP_j} = \frac{\sum_{i=1}^{n}{w_{i_j}}{x_i}}{\sum_{i=1}^{n}{w_{i_j}}} \]

w\textsubscript{ij} is the score of emotion term \emph{j} in sample \emph{i}, x\textsubscript{i} is the CoreGRID score vector for sample \emph{i} and \emph{n}  is the number of samples. We conducted a Ward method-based hierarchical clustering using the Euclidean distance metric \cite{RN25, RN646}. The hierarchical organisation illustrated in Figure \ref{fig:Figure4_CPMClusters} depicts the literature's classical definitions of positive and negative emotions \cite{RN190}. Further, it can be explained by the CPM's intrinsic (un)pleasantness, goal obstructiveness, and goal relevance \cite{RN97}. Lower-level branching can also be identified as eight meaningful sub-clusters: anger (hate, disgust, anger), fear, sadness (disappointment, sadness), embarrassment (shame, regret, guilt, contempt), happiness (pleasure, joy, amusement, pride), satisfaction (admiration, contentment), serenity (relief, interest) and affection (compassion, love). 

In our negative cluster in the CPM space, we found anger, fear, sadness, and embarrassment as the main sub-levels. These hierarchically organised clusters demonstrate the high arousal activation of anger and fear against low-arousal emotions such as sadness \cite{RN704}. Also, shame, regret, and guilt clustered in terms of embarrassment suggest the clustering of low power and negative valence emotions, as depicted in GEW \cite{RN190}. However, contempt, which is clustered in the higher-level branch of the embarrassment cluster, is defined as a negative valenced and high-power in GEW \cite{RN190}. But contempt clustered here with shame and guilt potentially due to the complex nature of these emotions as social or self-conscious emotions \cite{RN249}. In our positive cluster, we found happiness, satisfaction, serenity, and affection as the main sub-levels. The happiness cluster consists of high power and positive valence emotions agreeing to the independent research \cite{RN97, RN295}. Similarly, the hierarchical organisation of the satisfaction cluster agrees with GEW's low power and positive valenced emotions \cite{RN190}. However, a separate serenity cluster is observed with interest and relief, which are positive valence but opposed poled with power dimension regarding GEW \cite{RN190}. This observation contrasts with the previous findings where interest and relief were observed in two different clusters in CPM space \cite{RN67}. Our observation may be possibly due to the high immersion and pleasantness in selected VR game experiences. The hierarchical organisation of affection cluster in the CPM space agrees with the low power and positive valence dimensions, including attachment-related emotions such as compassion and love \cite{RN230, RN206}. 

The vector values of the weighted average CPM profile (CP\textsubscript{j}) of each discrete emotion term are presented in Supplementary Figure S2. Accordingly, the intensity of different CoreGRID items varied in response to different discrete emotions or combinations of them. This illustrates the impacting factors for each emotion and supports the previous clustering results.

Overall, our results' validity and meaningful clustering for a wide range of emotions show that VR is a powerful tool for emotion research, offering a more realistic environment than traditional methods such as films \cite{RN97, RN295}.   

\subsection{Labeling the Emotional Dimensions}
Emotion terms are differentiated by various dimensional models of emotion. While the basic valence and arousal dimensional model is generally recognised, there are complications in rationalising some emotions \cite{RN581}. For example, intense anger is considered high arousal, and intense sadness is considered low arousal \cite{RN189}. Given that, the basic dimensional model cannot distinguish between intense and less intense sadness due to low arousal in both cases. Therefore, the current scope of the basic dimensional model creates limitations in differentiating certain emotions. 

To explore the latent dimensions and address our second research question, we conducted an exploratory factor analysis on the 51 CoreGRID items, examining dimensions from a componential perspective. We performed a factor analysis followed by a Varimax rotation on the z-score normalised 51 CoreGRID items \cite{RN572} and identified five factors based on a scree plot. We found five meaningful factors explaining 43.84\% of the total variance, capturing aspects related to body changes and expressions associated with suddenness and novelty (14.20\%), valence (10.89\%), agency (6.94\%), novelty (5.42\%), and norms (6.40\%). Our results align with the individual research \cite{RN582, RN581, RN295, RN572}, suggesting more than the commonly considered two dimensions of valence and arousal. As an example, the full CPM data-driven study by \cite{RN295} identified action tendency, (un)pleasantness, novelty, norms, arousal, and goal relevance as dimensional representations of emotions. Some of the previous works have used a few components \cite{RN34}, emotional terms, and facial expressions \cite{RN230, RN708, RN709}, so the results were limited to a few dimensions. Nevertheless, our assumption of full CPM with CoreGRID items expands the dimensions that can differentiate among emotion words. Moreover, a multi-componential architecture of emotion aligns broadly with our five-factor analysis.

\subsection{Role of Facial EMG in the Emotional Experience}
\begin{figure*}
    \centering
    \includegraphics[width=0.9\linewidth]{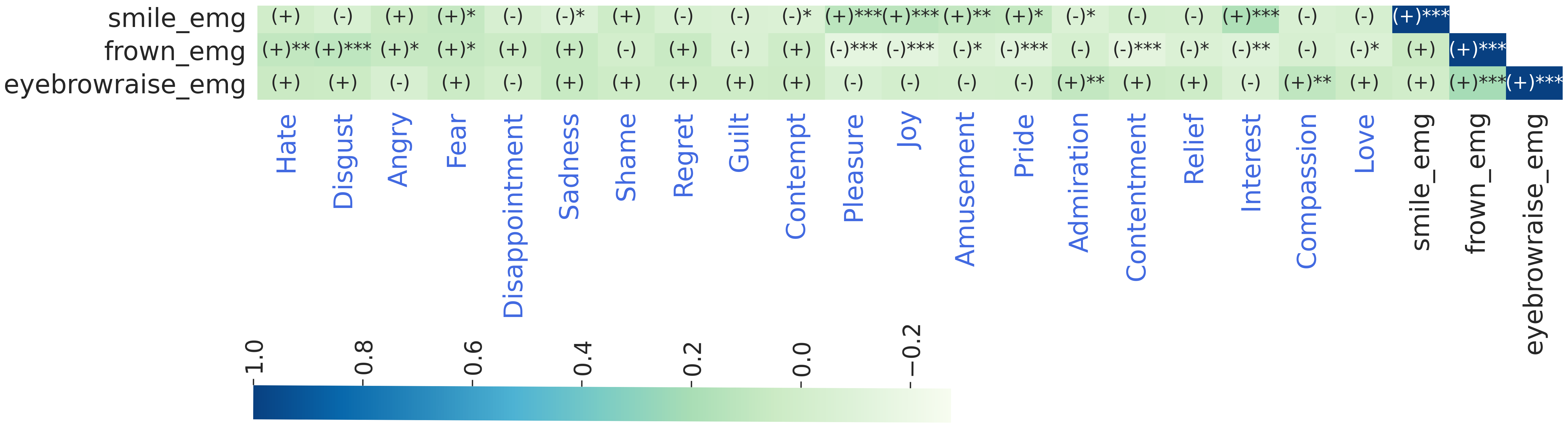}
    \caption{The correlation matrix displays the average intensity of three facial EMG expressions (\emph{``smile\_emg''}, \emph{``frown\_emg''}, \emph{``eyebrowraise\_emg''}) in black font and the GEW emotions in blue font. Positive correlations are indicated by a (+) symbol, and a (-) symbol indicates negative correlations. Asterisks indicate the statistical significance of the results at a p-value of: *p\textless0.05, **p\textless0.01, ***p\textless0.001.}
    \label{fig:Figure5_EMGEmo}
\end{figure*}

Facial expressions are a prominent form of non-verbal communication of emotions \cite{RN669, RN678}. However, when face monitoring is not feasible, facial EMG has been used to measure muscle movements using surface EMG sensors \cite{RN696}. Therefore, we investigated the power of facial EMG in differentiating self-reported emotional (20 emotion terms from GEW) experiences to gain insights into the role of CPM expression components in emotions and their correlation. We uploaded our calibration data (neutral, max smile, max frown, max eyebrow raises) and gameplay data to SuperVision to get facial EMG expressions (smile, frown, eyebrow raise) and their intensities. Figure \ref{fig:Figure5_EMGEmo} shows the Spearman correlation of three facial EMG expressions with 20 emotions.  

The results showed a significant positive correlation of \emph{``smile\_emg''} with interest, joy, pleasure (p\textless0.001), amusement (p\textless0.01), and pride, fear (p\textless0.05), indicating activation of zygomaticus muscles in pleasant scenarios except for fear \cite{RN664, RN705, RN479, RN678}. However, the positive correlation of EMG between fear, shame, hate, and anger may be due to lip tightener, lip stretcher, and jaw drop \cite{RN667}, which are related to mouth areas. A significant negative correlation was observed with admiration, sadness, and contempt  (p\textless0.05), possibly due to lower arousal \cite{RN190} and complexity \cite{RN249, RN207} of these emotions, leading to lower activation in the zygomaticus muscle. The \emph{``frown\_emg''} was positively correlated with most of the negative emotions disgust (p\textless0.001), hate (p\textless0.01), fear, and anger (p\textless0.05). Further, the positive emotions except admiration and compassion were significantly negatively correlated with \emph{``frown\_emg''}. The high activation of corrugator supercilli facial muscles in negative experiences can explain the observation \cite{RN705, RN696, RN7}.To further note, the significant correlation between the corrugator or \emph{``frown\_emg''} and positive/negative emotions can be attributed to its positioning below the frontalis muscle, which is involved in facial expressions such as frowning and eyebrow-raising \cite{RN696}.

The pattern of \emph{``eyebrowraise\_emg''} with emotions was ambiguous, given that it showed a significant positive correlation with \emph{``frown\_emg''} (p\textless0.001), admiration, and compassion (p\textless0.01). The positive correlations with admiration and compassion, typically associated with positive emotions, suggest that \emph{``eyebrowraise\_emg''} may not be a definitive indicator of positive emotions. The significant negative correlation of \emph{``eyebrowraise\_emg''} with interest and pleasure (p\textless0.01) may be due to the high arousal of these emotions showing lower activations of frontalis EMG \cite{RN664}. Although most were non-significant, the positive correlation between \emph{``eyebrowraise\_emg''} and emotions, except for seven, suggest ambiguous valence \cite{RN694} and the role of novelty \cite{RN581}, more like in surprise elicitation. 

\begin{figure}[htbp]
    \centering
    \includegraphics[width=0.95\linewidth]{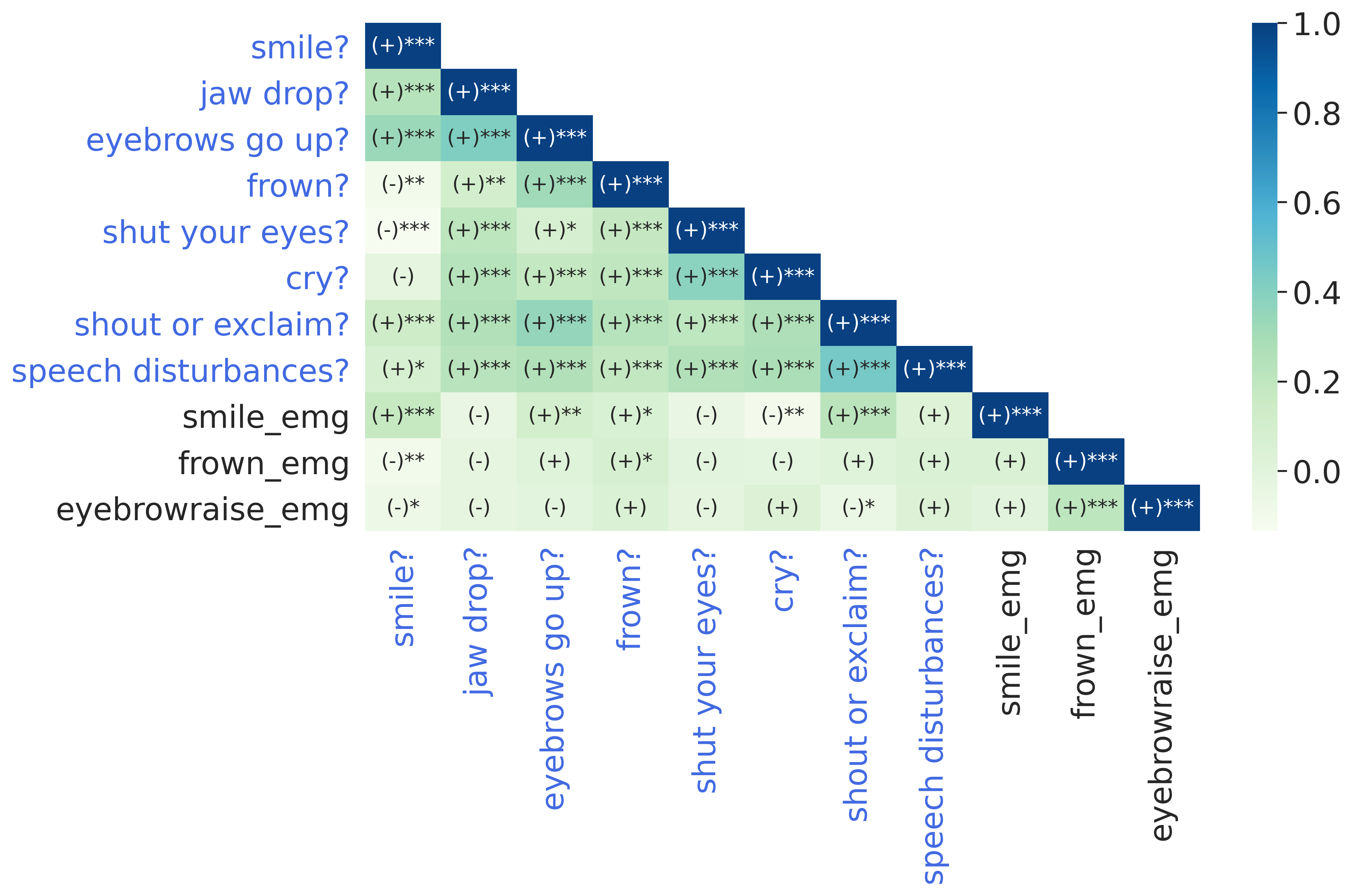}
    \caption{The correlation matrix presents the average intensity of three facial EMG expressions (\emph{``smile\_emg''}, \emph{``frown\_emg''}, \emph{``eyebrowraise\_emg''}) in black font and the CoreGRID expression component in blue font. Positive correlations are denoted by (+), while negative correlations are represented by (-). Asterisks indicate the statistical significance of the results at a p-value of: *p\textless0.05, **p\textless0.01, ***p\textless0.001.}
    \label{fig:Figure6_EMGGRID}
\end{figure}

Similarly, we did a Spearman correlation but with the eight CoreGRID expression component items and three facial EMG activations. As given in Figure \ref{fig:Figure6_EMGGRID}, a positive correlation of \emph{``smile\_emg''} with \emph{``smile?''}, \emph{``shout or exclaim?''} (p\textless0.001), \emph{``eyebrows go up?''} (p\textless0.01), and \emph{``frown?''} (p\textless0.05) were visible. The \emph{``smile\_emg''} correlation between \emph{``smile?''}, \emph{``shout or exclaim?''} was expected due to the zygomaticus muscle movement caused by an actual smile \cite{RN664, RN662} and involuntary speech or orofacial movements \cite{RN678}, respectively, where SuperVision cannot differentiate between two expressions linked to similar muscle areas. The correlation of \emph{``smile\_emg''} with self-reported \emph{``frown?''} may be due to the activations in the cheek region, such as nose wrinkling, upper lip raising, lip corner depression, and lips tightening in unpleasant outcomes \cite{RN479}. In contrast, \emph{``smile\_emg''} was negatively correlated with \emph{``cry?''} (p\textless0.01), which is consistent with the lower activation of the zygomaticus muscle previously observed in other studies \cite{RN707}. Given the differences in individual experiment settings, insights from SuperVision may be more general, so a customised model may improve interpretations.

As the next observation, \emph{``frown\_emg''} showed a positive correlation with \emph{``frown?''} (p\textless0.05) and a negative correlation with \emph{``smile?''} (p\textless0.01). This may be possibly due to the high and opposite activation of corrugator supercilli muscles \cite{RN613, RN665} respectively or due to the involvement of corrugator muscle in both negative and non-negative facial expressions \cite{RN696}. This finding aligns with the obstructive and conducive appraisal in the brow region \cite{RN479}. 

The \emph{``eyebrowraise\_emg''} showed a positive correlation with \emph{``frown\_emg''} (p\textless0.001) and negative correlations with \emph{``smile?''} (p\textless0.05). The negative correlations suggest that raised eyebrows are less likely to occur with smiling, possibly indicating a more subdued emotional state or a lack of intensity in the emotional response. However, the correlation between EMG and self-reported eyebrow activations was negative and non-significant, perhaps due to certain oversights in the SuperVision application or individuals’ perception of eyebrows raising is not necessarily reliable. 

In any case, the results demonstrate the possibility of facial EMG in monitoring facial expressions to detect emotions. Further, it emphasises the role of the expression component in emotion formation after event appraisal. 

\subsection{Modelling using Machine Learning}
To test for any profound relationships of emotions with full CPM, including signals, we used ML classifiers. This section focuses specifically on our third research question, which seeks to determine the extent to which each component of the CPM contributes to the identification of discrete emotions. For that, we first preprocessed the data and applied digital signal processing. EMG signals are typically sampled at a rate of 1000 Hz to capture and preserve the signal's full range of features and nuances \cite{RN669}. Therefore, we resampled all the signals to 1000 Hz to ensure that the signals are evenly sampled in time and allow for accurate comparison and analysis of the signals. The BVP signal was up-sampled from 64 Hz to 1000 Hz, and a median filter was applied to reduce the noise \cite{RN657}. Both skin temperature and EDA up sampled from 4 Hz to 1000 Hz. Then applied, a Savitzky-Golay (SG) filter \cite{RN659} for skin temperature (\emph{window\_length}=9 and \emph{polyorder}=5) and EDA (\emph{window\_length}=11 and \emph{polyorder}=5) \cite{RN700} for smoothing the signal. The respiration signal was up-sampled from 256 Hz to 1000 Hz and applied an SG filter (\emph{window\_length}=11 and \emph{polyorder}=5) \cite{RN700}. Seven EMG amplitude channels (right and left frontalis, right and left orbicularis, right and left zygomaticus, corrugator) were downsampled to 1000 Hz from 2000 Hz. 

For feature extraction, we employed Discrete Wavelet Transformations (DWT). DWT is beneficial in this context because it allows for localisation in both the frequency and time domains and provides a multi-resolution analysis at different frequencies or scales. Through experimentation, we found that the Daubechies wavelet family and five levels of frequency decomposition yielded the best results. We used the coefficients' mean, variance, and median resulting from the five decomposition levels as features. Accordingly, we used 51 CoreGRID items, three statistical features of DWT of BVP (15 features), skin temperature (15 features), EDA (15 features), respiration (15 features), and seven EMG channels (105 features). Apart from that, we fed the average intensities of smile, frown, and eyebrow raise retrieved from the SuperVision application, resulting in 219 input features.

We trained ML classifiers treating the CoreGRID items and signal features as input representing the CPM interpretations. We used each discrete emotion's high and low values as split by the first quartile output. Due to the imbalance between high and low values, we selected the first quartile as the threshold instead of the median. We within-subject z-normalised the input features and sampled the training dataset using the Synthetic Minority Oversampling Technique (SMOTE) \cite{RN503} to reduce the imbalance in the dataset. For better generalisability, we report the LOSO cross-validation results, taking the average after all iterations. 

\subsubsection{Interpretation of Emotional Experience through the CoreGRID and Physiological Changes}
To understand the emotional experience through the CoreGRID items and physiological changes, we trained several commonly used ML classifiers such as Support Vector Machine (SVM), Random Forest (RF), eXtreme Gradient Boosting (XGB) and Light Gradient Boosting Machine (LGBM). We followed the process outlined in Section 4.5 and evaluated the performance of our model using LOSO cross-validation, repeating the process. Although we trained several classification models, we found that the LGBM had the best performance, so we presented the results from that model.

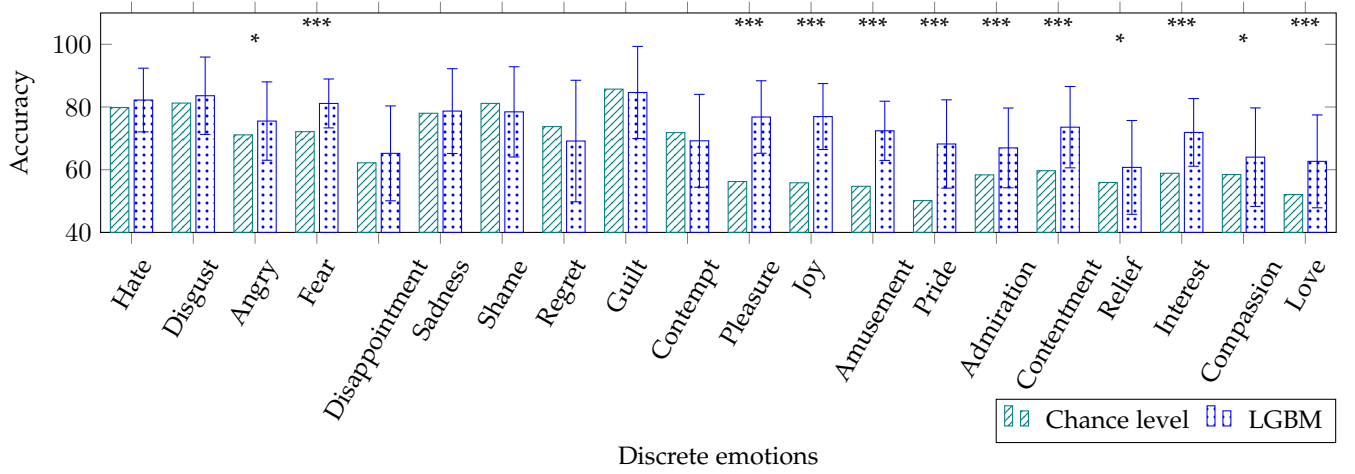
\begin{figure*}[htbp]
\centering
\begin{tikzpicture}
    \begin{axis}[
        height=4.5cm,
        width=18cm,
        ymin=40,
        ymax=110,
        extra y tick labels={},
        ybar,
        bar width=7pt,
        xtick distance=0.5,
        ylabel=Accuracy,
        xlabel=Discrete emotions,
        enlarge x limits={abs=0.5},
        scaled ticks=false,
        xtick={1,2,3,4,5,6,7,8,9,10,11,12,13,14,15,16,17,18,19,20},
        xticklabels={
        {Hate},{Disgust},{Angry},{Fear},{Disappointment},{Sadness},{Shame},{Regret},{Guilt},   {Contempt},{Pleasure},{Joy},{Amusement},{Pride},{Admiration},{Contentment},{Relief},{Interest},{Compassion},{Love}},
        xticklabel style={align=center,rotate=60},
        legend style={at={(1,-0.96)},anchor=south east,column sep=1ex},
        legend columns = 2,
        legend cell align={left},
    ]
    
        \addplot [teal, pattern=north east lines, pattern color=teal] coordinates {
            (1,79.84) 
            (2,81.26) 
            (3,71.12) 
            (4,72.14) 
            (5,62.21) 
            (6,78.01) 
            (7,81.16) 
            (8,73.76) 
            (9,85.71) 
            (10,71.83) 
            (11,56.23) 
            (12,55.83) 
            (13,54.71) 
            (14,50.15) 
            (15,58.36) 
            (16,59.68) 
            (17,55.93) 
            (18,58.87) 
            (19,58.46) 
            (20,52.08) 
            };

        \addplot+ [error bars/.cd,y dir=both,y explicit] 
        [blue, pattern=dots, pattern color=blue]
        coordinates {
            (1,82.23) +- (0,10.16) 
            (2,83.60) +- (0,12.34) 
            (3,75.52) +- (0,12.50) 
            (4,81.14) +- (0,7.80) 
            (5,65.22) +- (0,15.13) 
            (6,78.70) +- (0,13.52) 
            (7,78.46) +- (0,14.39) 
            (8,69.15) +- (0,19.38) 
            (9,84.64) +- (0,14.69) 
            (10,69.22) +- (0,14.82) 
            (11,76.81) +- (0,11.56) 
            (12,76.96) +- (0,10.52) 
            (13,72.41) +- (0,9.44) 
            (14,68.21) +- (0,14.08) 
            (15,66.96) +- (0,12.72) 
            (16,73.57) +- (0,12.97) 
            (17,60.74) +- (0,14.96) 
            (18,71.90) +- (0,10.80) 
            (19,64.01) +- (0,15.71) 
            (20,62.67) +- (0,14.81) 
            };

        \draw (axis cs:0,145) ++ (3.80,5pt) -- ++(10pt,0pt);
        \node[anchor=south] at (axis cs:3,95) {*};
        \draw (axis cs:0,145) ++ (4.80,5pt) -- ++(10pt,0pt);
        \node[anchor=south] at (axis cs:4,100) {***};
        \draw (axis cs:0,145) ++ (11.80,5pt) -- ++(10pt,0pt);
        \node[anchor=south] at (axis cs:11,100) {***};
        \draw (axis cs:0,148) ++ (12.80,5pt) -- ++(10pt,0pt);
        \node[anchor=south] at (axis cs:12,100) {***};
        \draw (axis cs:0,145) ++ (13.80,5pt) -- ++(10pt,0pt);
        \node[anchor=south] at (axis cs:13,100) {***};
        \draw (axis cs:0,145) ++ (14.80,5pt) -- ++(10pt,0pt);
        \node[anchor=south] at (axis cs:14,100) {***};
        \draw (axis cs:0,145) ++ (15.80,5pt) -- ++(10pt,0pt);
        \node[anchor=south] at (axis cs:15,100) {***};
        \draw (axis cs:0,148) ++ (16.80,5pt) -- ++(10pt,0pt);
        \node[anchor=south] at (axis cs:16,100) {***};
        \draw (axis cs:0,148) ++ (17.80,5pt) -- ++(10pt,0pt);
        \node[anchor=south] at (axis cs:17,95) {*};
        \draw (axis cs:0,145) ++ (18.80,5pt) -- ++(10pt,0pt);
        \node[anchor=south] at (axis cs:18,100) {***};
        \draw (axis cs:0,148) ++ (19.80,5pt) -- ++(10pt,0pt);
        \node[anchor=south] at (axis cs:19,95) {*};
        \draw (axis cs:0,145) ++ (20.80,5pt) -- ++(10pt,0pt);
        \node[anchor=south] at (axis cs:20,100) {***};
        \legend{Chance level, LGBM}
    \end{axis}
\end{tikzpicture}
\caption{Accuracy of the LGBM binary classifiers for differentiating emotions. Accuracy is compared with the chance level (majority class portion). Error bars represent the standard deviation. Asterisks indicate the significance of results at p-value: *p\textless0.05, **p\textless0.01, ***p\textless0.001. }
\label{fig:Figure6_fullpredictgew}
\end{figure*}

We initially assessed model performance using CoreGRID items alone and in combination with physiological measures. The combined approach showed an average increase of 1.61\%, ranging from 0.22\% to 4.48\%. Therefore, the results of training the LGBM classifier using both measures for each emotion against the chance level (majority class prediction) are shown in Figure \ref{fig:Figure6_fullpredictgew}. In our analysis, we generated the baseline using both the chance level and a nonparametric chance distribution through permutations. However, since both baselines were nearly similar, we opted to use the chance level as the baseline for all of our analyses. In all cases except four (shame, regret, guilt, and contempt), the model exceeds the chance level. However, according to the one-sample t-tests, accuracy was significantly higher (p\textless0.001) for interest, amusement, pride, joy, pleasure, contentment, admiration, love, and fear. Accuracy of relief, compassion, and anger are moderately significant (p\textless0.05). The lower performance for the remaining emotions can be attributed to the skewed distribution of the emotion ratings, resulting in fewer representative examples of the underrepresented class, and making it more difficult for the model to identify distinguishing patterns. For example, guilt, disgust, and shame are heavily skewed towards the lower end, possibly due to the complexity of these emotions \cite{RN207}. As VR games are primarily designed for entertainment, it is not easy to find games that effectively elicit these emotions. Additionally, the margin of improvement for some significant models is smaller due to the same cause. Overall, the effectiveness of selected physiological measures and CoreGRID items in differentiating emotional features is depicted. 

\subsubsection{Role of each CPM Component and Physiological Changes in Differentiating Discrete Emotions}
To investigate the contribution of each CPM component and physiological measures, we trained ML models as previously described but using the relevant combination of features for each component. (For example, the appraisal component includes 17 appraisal CoreGRID items, and the physiology component includes 8 physiology CoreGRID items and BVP, skin temperature, EDA, and respiration). The results for 20 emotions and the statistical significance level for using either individual components or full CPM compared to chance level (majority class prediction) are given in Table \ref{tab:Table1ML}. In Table \ref{tab:Table2ML}, we provide the results when removing one component at a time and training with the remaining components. Using this approach, we can determine the additional information provided by each component in distinguishing discrete emotions.

\begin{table*}[htbp]
\vspace{-0.3cm}
\centering
\caption{Accuracy of the LGBM binary classifiers for the differentiation of each emotion using individual components. Asterisks indicate the significance of results at p-value: *p\textless0.05, **p\textless0.01, ***p\textless0.001. The chance level is used as the baseline. Appraisal, motivation, and feeling components include only the relevant CoreGRID items. The physiology component includes CoreGRID items and BVP, skin temperature, EDA, and respiration. The expression component contains CoreGRID items, EMG amplitude signals, and three facial expression values. Full CPM includes all CoreGRID items and signal data.}
\vspace{-0.3cm}
\begin{tabular}{l|r|r|r|r|r|r|r}
    & \textbf{Appraisal} & \textbf{Motivation} & \textbf{Expression} & \textbf{Physiology} & \textbf{Feeling} & \textbf{Full CPM} & \textbf{Baseline} \\ \hline
    \textbf{Hate}           & 77.99 & 80.25 & 72.11 & 68.56 & 72.73 & 82.23 & 79.84 \\
    \textbf{Disgust}        & 74.83 & 77.59 & 77.98 & 70.62 & 68.82 & 83.60 & 81.26 \\
    \textbf{Angry}          & 70.09 & 70.14 & 68.84 & 66.08 & 69.53 & 75.52* & 71.12 \\
    \textbf{Fear}           & 73.52 & 77.30** & 73.87 & 75.77* & 76.02** & 81.14*** & 72.14 \\
    \textbf{Disappointment} & 66.13 & 59.53 & 57.16 & 58.61 & 59.96 & 65.22 & 62.21 \\
    \textbf{Sadness}        & 70.45 & 68.16 & 73.91 & 68.48 & 69.29 & 78.70 & 78.01 \\
    \textbf{Shame}          & 70.97 & 70.62 & 75.05 & 70.24 & 66.58 & 78.46 & 81.16 \\
    \textbf{Regret}         & 72.10 & 65.79 & 61.01 & 61.95 & 66.48 & 69.15 & 73.76 \\
    \textbf{Guilt}          & 77.27 & 76.88 & 82.03 & 76.09 & 66.41 & 84.64 & 85.71 \\
    \textbf{Contempt}       & 67.55 & 68.53 & 61.86 & 63.52 & 61.32 & 69.22 & 71.83 \\
    \textbf{Pleasure}       & 73.64*** & 71.83*** & 74.34*** & 60.45* & 75.61*** & 76.81*** & 56.23 \\
    \textbf{Joy}            & 73.66*** & 71.40*** & 76.36*** & 61.66** & 75.47*** & 76.96*** & 55.83 \\
    \textbf{Amusement}      & 68.40*** & 70.43*** & 70.08*** & 56.14 & 69.79*** & 72.41*** & 54.71 \\
    \textbf{Pride}          & 69.22*** & 62.91*** & 64.83*** & 60.96*** & 68.52*** & 68.21*** & 50.15 \\
    \textbf{Admiration}     & 61.83 & 60.50 & 63.25** & 60.78 & 65.77*** & 66.96*** & 58.36 \\
    \textbf{Contentment}    & 69.89*** & 65.96** & 65.25** & 62.93 & 73.93*** & 73.57*** & 59.68 \\
    \textbf{Relief}         & 55.20 & 47.32 & 57.96 & 58.00 & 58.10 & 60.74* & 55.93 \\
    \textbf{Interest}       & 64.99** & 68.24*** & 67.44*** & 56.80 & 67.63*** & 71.90*** & 58.87 \\
    \textbf{Compassion}     & 58.83 & 60.84 & 52.46 & 56.70 & 59.95 & 64.01* & 58.46 \\
    \textbf{Love}           & 58.06** & 54.68 & 62.09*** & 58.21** & 58.95** & 62.67*** & 52.08 \\
\end{tabular}
\label{tab:Table1ML}
\end{table*}

Accordingly, the appraisal component was significant for amusement, pride, joy, pleasure, and contentment (p\textless0.001) and interest, love (p\textless0.01). This correlation between the appraisal component and most positive emotions may be due to the compatibility with the participant's goals, their ability to cope with the experience, and the intrinsic pleasantness \cite{RN97, RN111} of the VR experience. The motivation component showed significant performance for interest, amusement, pride, joy, pleasure (p\textless0.001), contentment, and fear (p\textless0.01). It appears that motivational goals such as information seeking, savouring, and wishful thinking \cite{RN111} have impacted the experience of positive emotion. In contrast, fear correlations represented avoidance action tendencies \cite{RN248}. Next, for the CPM expression component, first, we trained classifiers only using the self-reports and both self-reports and facial signal features. We reached the conclusion that models incorporating both inputs consistently outperformed unimodal models, with improvements ranging from 0.02\% to 12.77\%. Therefore, this paper only presents the results for these multimodal models. Accordingly, the expression component was significant for interest, amusement, pride, joy, pleasure, love (p\textless0.001), contentment, and admiration (p\textless0.01). This finding suggests that facial expressions, the most natural form of nonverbal communication \cite{RN669}, are effectively captured by facial EMG signals and CoreGRID items.

Like the expression component, a better performance (0.45\% - 8.27\% improvement) was observed for the multimodal physiological component with a combination of self-reports and signal features. In parallel with previous research, the physiology component showed ambiguous patterns for emotion differentiation \cite{RN206, RN590}. Consequently, emotions such as pride (p\textless0.001), joy, love (p\textless0.01), pleasure, and fear (p\textless0.05) showed significance. Although fear showed similar behaviour, the remaining significant observations diverged from those seen in a previous study \cite{RN25}. In that study, it was observed that only the expression and feeling components held significance for the emotions of joy, love, and satisfaction. One potential explanation could be that our study, which employed both subjective and objective measures, possessed more informative features.

Lastly, the feeling component is significant for interest, amusement, pride, joy, pleasure, contentment, admiration (p\textless0.001), love, and fear (p\textless0.01). However, the feeling component, traditionally the focus of previous research, demonstrated minimal performance in predicting most negative emotions. This less prediction capacity for negative emotions can be partially explained by the skewness of the emotion ratings towards the lower end because VR experiences were often perceived as pleasant even when they involved challenging or fearful content \cite{RN67}. Overall, similar to using full CPM, using single components demonstrates better distinction powers for positive emotions in our experiment.   

\begin{table*}[t]
\centering
\caption{Accuracy of the LGBM binary classifiers for the differentiation of each emotion when removing one component at a time. Asterisks indicate the significance of results at p-value: *p\textless0.05, **p\textless0.01, ***p\textless0.001. The full CPM is used as the baseline. For classification without appraisal, motivation, and feeling components, only the relevant CoreGRID items were removed. For classification excluding the physiology component, CoreGRID physiology questions, BVP, skin temperature, EDA, and respiration were removed. For classification without the expression component, CoreGRID expression questions, EMG amplitude signals, and three facial expression values were removed. Full CPM includes all CoreGRID items and signal data. (WO: Without)}
\vspace{-0.3cm}
\begin{tabular}{l|r|r|r|r|r|r}
    & \textbf{WO Appraisal} & \textbf{WO Motivation} & \textbf{WO Expression} & \textbf{WO Physiology} & 
    \textbf{WO Feeling} & \textbf{Full CPM} \\ \hline
    \textbf{Hate} & 81.34 & 81.27 & 82.12 & 81.49 & 82.01 & 82.23 \\ 
    \textbf{Disgust} & 83.76 & 80.37** & 81.78* & 84.32 & 83.92 & 83.60 \\
    \textbf{Angry} & 75.03 & 72.87* & 76.15 & 75.42 & 74.34 & 75.52* \\
    \textbf{Fear} & 81.16 & 79.55 & 81.12 & 80.63 & 80.26 & 81.14*** \\
    \textbf{Disappointment} & 65.62 & 65.77 & 64.00 & 63.32 & 67.45 & 65.22 \\
    \textbf{Sadness} & 76.16** & 76.47* & 75.86* & 77.80 & 76.23 & 78.70 \\
    \textbf{Shame} & 79.05 & 77.94 & 75.79* & 78.09 & 77.83 & 78.46 \\
    \textbf{Regret} & 70.17 & 69.94 & 72.4** & 69.64 & 68.76 & 69.15 \\
    \textbf{Guilt} & 84.16 & 83.12 & 83.76 & 84.23 & 85.28 & 84.64 \\
    \textbf{Contempt} & 70.83 & 65.33 & 70.67 & 69.06 & 68.96 & 69.22 \\
    \textbf{Pleasure} & 76.06 & 77.70 & 75.28 & 77.60 & 75.61 & 76.81*** \\
    \textbf{Joy} & 78.37 & 78.09 & 77.98 & 77.16 & 77.51 & 76.96*** \\
    \textbf{Amusement} & 71.30 & 70.43 & 70.95 & 71.42 & 72.68 & 72.41*** \\
    \textbf{Pride} & 66.35 & 68.42 & 68.20 & 69.12 & 68.10 & 68.21*** \\
    \textbf{Admiration} & 66.38 & 66.10 & 67.91 & 65.03 & 66.31 & 66.96*** \\
    \textbf{Contentment} & 72.28 & 73.19 & 72.48 & 72.88 & 69.95* & 73.57*** \\
    \textbf{Relief} & 61.45 & 59.65 & 59.53 & 55.40* & 59.98 & 60.74* \\
    \textbf{Interest} & 71.49 & 69.65 & 69.55 & 69.88* & 70.38 & 71.90*** \\
    \textbf{Compassion} & 60.14* & 62.50 & 62.37 & 62.16 & 61.83 & 64.01* \\
    \textbf{Love} & 59.38 & 61.33 & 59.88 & 60.34 & 61.46 & 62.67*** \\
\end{tabular}
\label{tab:Table2ML}
\end{table*}

Next, we compared the results of training ML models using single components (Table \ref{tab:Table1ML}) with results from excluding one component (Table \ref{tab:Table2ML}). To better understand the role and importance of each component, in this experiment, we compared the accuracy of each model with full CPM as the baseline (Table \ref{tab:Table2ML}), as our objective here is to show the relative importance of each component in the full CPM model. We expect that removing more informative components will result in higher reductions in the prediction power.

Table \ref{tab:Table1ML} showed that emotions like pleasure, joy, pride, fear, amusement, and admiration were significant when using single components or with the full CPM. However, in Table \ref{tab:Table2ML}, removing each component did not lead to any significant difference when comparing these emotions (pleasure, joy, pride, fear, amusement, and admiration) with the full CPM results. This suggests that each component can complement specific emotion differentiation, and the information from removed components is already encoded in other components.

As shown in Table \ref{tab:Table1ML}, when comparing accuracy with the chance level for emotions like disgust, sadness, and shame, no scenario was significant. However, removing different combinations of motivation, expression, and appraisal components resulted in a significant accuracy reduction compared to the full CPM performance (Table \ref{tab:Table2ML}). This implies that although these components may not directly impact the results, they provide complementary information not provided by other components or the full CPM.

For interest, the full CPM (p\textless0.001) and single components, except physiology, were significant. However, a significant performance reduction was observed when using all components except physiology, suggesting that even though it may not be important as a single component, it encodes relevant complementary information for interest. Similarly, for contentment, including only the feeling component (p\textless0.001) and removing the feeling component (p\textless0.01) both showed significant performance. This indicates that the feeling component provides significant details for contentment.

A different observation was made for regret, where neither the single components nor the full CPM were significant, as shown in Table \ref{tab:Table1ML}. However, removing the expression component (p\textless0.01) in Table \ref{tab:Table2ML} led to an accuracy improvement. This implies that the information encoded in the expression component might be just redundant. On the other hand, for the emotions of hate, disappointment, guilt, and contempt, no model was significant in both tables. In summary, this overall comparison suggests that while some components have complementary counterparts, others contain very specific or complementary information for each emotion.

As the last analysis, we report the feature importance scores for the LGBM model while using the full CPM for emotion prediction. Supplementary material Figures S3-S7 illustrate the scores for each component while trained using the full CPM. Accordingly, a clear differentiation can be observed among the clusters depicted in Figure \ref{fig:Figure4_CPMClusters}.

Overall, the model that includes all CPM self-reports and DWT signal features performs well in distinguishing emotions. Our results reveal that each component of the CPM may have a specific role in emotion differentiation. Additionally, considering only one component of the CPM may limit our understanding of emotion formation. These findings suggest that a complete understanding of emotion formation may require considering the interplay between all five components of the CPM.

The accuracy improvements observed when using both subjective and objective measures highlight the value of incorporating objective measures in addition to subjective evaluations, particularly for components of emotion that individuals may not be aware of. This suggests that relying solely on subjective evaluations may not provide a complete understanding of emotional states.

\section{Conclusion}
Our manuscript aimed to study CPM as a framework to examine the interconnected components and sub-processes that contribute to the formation of emotions. We operationalised a data collection using 27 interactive VR games and measured the subjective ratings and physiological and facial signals from 39 participants. This study makes several significant contributions to the field of affective computing and emotion recognition: (i) Using VR as an immersive environment to induce emotions as a more ecologically valid setup, (ii) It provides insights into the underlying mechanisms that link CPM and emotional experience using a realistic emotion induction paradigm without assuming any pre-labels for emotional experience. This allows for a better understanding of how these factors are related in a naturalistic setting and takes into account individual differences, (iii) Provides a comprehensive analysis of the underlying dimensions that describe the emotional experience, (iv) The study examines the contribution of different CPM components and modalities to emotional experience, providing a nuanced and detailed understanding of their respective roles, and (v) The study employs a larger sample size and generalised ML models to examine emotion formation, which allows for more accurate and reliable findings.

Our study found that the CPM provides a good framework for understanding emotion formation when emotions are induced through interactive VR games, and it can help differentiate between discrete emotions. We chose VR games due to their immersive nature, providing a richer emotional experience compared to still images or video clips. Thus, our focus in this manuscript is solely on understanding emotions rather than analysing the impact on VR gameplay. Each component of the CPM framework plays a specific role in capturing these emotions, and a combination of self-report and objective measures is effective in understanding their underlying processes. We identified five latent dimensions that underlie emotional experiences induced through interactive VR games, which may be universal to the human experience of emotions. These findings have implications for the design of systems that aim to recognise and respond to emotions, such as in healthcare, education, and gaming. Further research is needed to fully understand the complex processes underlying emotion formation and develop more effective methods for capturing and analysing emotions. 

It's important to acknowledge several limitations in this study. Firstly, the imbalanced distribution of emotions is likely influenced by the overall pleasantness and immersive nature of the VR experience. Additionally, we collected various other signals that were not utilised in our current analysis. Furthermore, increasing the sample size could enhance the robustness of our interpretations. Moreover, we recognise the challenge of accurately capturing and reporting facial expressions lasting mere milliseconds within a 3-minute period, along with participants' difficulty in recalling and reporting these subtle expressions comprehensively. In future research, investigating the impact of VR games on emotions, exploring the dynamic nature of emotional responses, and understanding brain processes could offer valuable insights. Moving forward, we recognise the importance of addressing these limitations in our future research works. 

\ifCLASSOPTIONcaptionsoff
  \newpage
\fi

\bibliographystyle{IEEEtran}
\bibliography{main}

\begin{thebibliography}{10}
\providecommand{\url}[1]{#1}
\csname url@samestyle\endcsname
\providecommand{\newblock}{\relax}
\providecommand{\bibinfo}[2]{#2}
\providecommand{\BIBentrySTDinterwordspacing}{\spaceskip=0pt\relax}
\providecommand{\BIBentryALTinterwordstretchfactor}{4}
\providecommand{\BIBentryALTinterwordspacing}{\spaceskip=\fontdimen2\font plus
\BIBentryALTinterwordstretchfactor\fontdimen3\font minus \fontdimen4\font\relax}
\providecommand{\BIBforeignlanguage}[2]{{%
\expandafter\ifx\csname l@#1\endcsname\relax
\typeout{** WARNING: IEEEtran.bst: No hyphenation pattern has been}%
\typeout{** loaded for the language `#1'. Using the pattern for}%
\typeout{** the default language instead.}%
\else
\language=\csname l@#1\endcsname
\fi
#2}}
\providecommand{\BIBdecl}{\relax}
\BIBdecl

\bibitem{RN20}
K.~R. Scherer, ``Emotions are emergent processes: they require a dynamic computational architecture,'' \emph{Philosophical Transactions of the Royal Society B: Biological Sciences}, vol. 364, no. 1535, pp. 3459--3474, 2009.

\bibitem{RN632}
R.~W. Picard, \emph{Affective computing}.\hskip 1em plus 0.5em minus 0.4em\relax Cambridge: MIT press, 2000.

\bibitem{RN676}
H.~Kober, L.~F. Barrett, J.~Joseph, E.~Bliss-Moreau, K.~Lindquist, and T.~D. Wager, ``Functional grouping and cortical--subcortical interactions in emotion: a meta-analysis of neuroimaging studies,'' \emph{Neuroimage}, vol.~42, no.~2, pp. 998--1031, 2008.

\bibitem{RN645}
K.~A. Lindquist, T.~D. Wager, H.~Kober, E.~Bliss-Moreau, and L.~F. Barrett, ``The brain basis of emotion: a meta-analytic review,'' \emph{Behavioral and brain sciences}, vol.~35, no.~3, pp. 121--143, 2012.

\bibitem{RN182}
J.~J.~R. Fontaine, \emph{Dimensional, basic emotion, and componential approaches to meaning in psychological emotion research1}.\hskip 1em plus 0.5em minus 0.4em\relax Oxford: Oxford University Press, 2013.

\bibitem{RN79}
D.~Grandjean, D.~Sander, and K.~Scherer, ``Conscious emotional experience emerges as a function of multilevel, appraisal-driven response synchronization,'' \emph{Consciousness and cognition}, vol.~17, pp. 484--95, 2008.

\bibitem{RN22}
E.~Harmon-Jones, C.~Harmon-Jones, and E.~Summerell, ``On the importance of both dimensional and discrete models of emotion,'' \emph{Behavioral sciences}, vol.~7, no.~4, p.~66, 2017.

\bibitem{RN473}
T.~Song, W.~Zheng, C.~Lu, Y.~Zong, X.~Zhang, and Z.~Cui, ``Mped: A multi-modal physiological emotion database for discrete emotion recognition,'' \emph{IEEE Access}, vol.~7, pp. 12\,177--12\,191, 2019.

\bibitem{RN23}
P.~J. Bota, C.~Wang, A.~L.~N. Fred, and H.~P.~D. Silva, ``A review, current challenges, and future possibilities on emotion recognition using machine learning and physiological signals,'' \emph{IEEE Access}, vol.~7, pp. 140\,990--141\,020, 2019.

\bibitem{RN73}
J.~A.~M. Correa, M.~K. Abadi, N.~Sebe, and I.~Patras, ``Amigos: A dataset for affect, personality and mood research on individuals and groups,'' \emph{IEEE Transactions on Affective Computing}, pp. 1--1, 2018.

\bibitem{RN28}
S.~Koelstra, C.~Muhl, M.~Soleymani, J.~Lee, A.~Yazdani, T.~Ebrahimi, T.~Pun, A.~Nijholt, and I.~Patras, ``Deap: A database for emotion analysis ;using physiological signals,'' \emph{IEEE Transactions on Affective Computing}, vol.~3, no.~1, pp. 18--31, 2012.

\bibitem{RN295}
G.~Mohammadi and P.~Vuilleumier, ``A multi-componential approach to emotion recognition and the effect of personality,'' \emph{IEEE Transactions on Affective Computing}, pp. 1--1, 2020.

\bibitem{RN471}
S.~Ojha, J.~Vitale, and M.-A. Williams, ``Computational emotion models: A thematic review,'' \emph{International Journal of Social Robotics}, 2020.

\bibitem{RN607}
K.~R. Scherer, ``Towards a prediction and data driven computational process model of emotion,'' \emph{IEEE Transactions on Affective Computing}, vol.~12, no.~2, pp. 279--292, 2021.

\bibitem{RN247}
P.~Ekman, \emph{Basic Emotions}, 2005, pp. 45--60.

\bibitem{RN131}
R.~Plutchik, ``A psychoevolutionary theory of emotions,'' vol.~21, no. 4-5, pp. 529--553, 1982.

\bibitem{RN67}
B.~Meuleman and D.~Rudrauf, ``Induction and profiling of strong multi-componential emotions in virtual reality,'' \emph{IEEE Transactions on Affective Computing}, vol.~PP, pp. 1--1, 2018.

\bibitem{RN21}
K.~R. Scherer, ``The dynamic architecture of emotion: Evidence for the component process model,'' \emph{Cognition and Emotion}, vol.~23, no.~7, pp. 1307--1351, 2009.

\bibitem{RN590}
M.~Q. Men{\'e}trey, G.~Mohammadi, J.~Leit{\~a}o, and P.~Vuilleumier, ``Emotion recognition in a multi-componential framework: the role of physiology,'' \emph{Frontiers in computer science}, vol.~4, p. 773256, 2022.

\bibitem{RN97}
B.~Meuleman and K.~R. Scherer, ``Nonlinear appraisal modeling: An application of machine learning to the study of emotion production,'' \emph{IEEE Transactions on Affective Computing}, vol.~4, no.~4, pp. 398--411, 2013.

\bibitem{RN650}
R.~Somarathna, T.~Bednarz, and G.~Mohammadi, ``Virtual reality for emotion elicitation - a review,'' \emph{IEEE Transactions on Affective Computing}, pp. 1--21, 2022.

\bibitem{RN649}
------, ``An exploratory analysis of interactive vr-based framework for multi-componential analysis of emotion,'' in \emph{2022 IEEE International Conference on Pervasive Computing and Communications Workshops and other Affiliated Events (PerCom Workshops)}, Conference Proceedings, pp. 353--358.

\bibitem{RN646}
------, ``Multi-componential analysis of emotions using virtual reality,'' in \emph{Proceedings of the 27th ACM Symposium on Virtual Reality Software and Technology}.\hskip 1em plus 0.5em minus 0.4em\relax Association for Computing Machinery, Conference Proceedings, p. Article 85.

\bibitem{RN701}
R.~Somarathna, A.~Quigley, and G.~Mohammadi, ``Multi-componential emotion recognition in vr using physiological signals,'' ser. AI 2022: Advances in Artificial Intelligence.\hskip 1em plus 0.5em minus 0.4em\relax Springer International Publishing, 2022, Conference Proceedings, pp. 599--613.

\bibitem{RN77}
J.~J.~R. Fontaine, K.~R. Scherer, and C.~Soriano, \emph{The why, the what, and the how of the GRID instrument1}.\hskip 1em plus 0.5em minus 0.4em\relax Oxford: Oxford University Press, 2013.

\bibitem{RN76}
K.~R. Scherer, J.~R.~F. Fontaine, and C.~Soriano, \emph{CoreGRID and MiniGRID: Development and validation of two short versions of the GRID instrument1}.\hskip 1em plus 0.5em minus 0.4em\relax Oxford: Oxford University Press, 2013.

\bibitem{marsella2010computational}
S.~Marsella, J.~Gratch, P.~Petta \emph{et~al.}, ``Computational models of emotion,'' \emph{A Blueprint for Affective Computing-A sourcebook and manual}, vol.~11, no.~1, pp. 21--46, 2010.

\bibitem{yongsatianchot2021computational}
N.~Yongsatianchot and S.~Marsella, ``Computational models of appraisal to understand the person-situation relation,'' in \emph{Measuring and Modeling Persons and Situations}.\hskip 1em plus 0.5em minus 0.4em\relax Elsevier, 2021, pp. 651--674.

\bibitem{RN69}
D.~Dupré, A.~Tcherkassof, and M.~Dubois, ``Emotions triggered by innovative products: A multi-componential approach of emotions for user experience tools,'' in \emph{2015 International Conference on Affective Computing and Intelligent Interaction (ACII)}, Conference Proceedings, pp. 772--777.

\bibitem{RN479}
K.~Gentsch, U.~Beermann, L.~Wu, S.~Trznadel, and K.~Scherer, ``Temporal unfolding of micro-valences in facial expression evoked by visual, auditory, and olfactory stimuli,'' \emph{Affective Science}, vol.~1, 2020.

\bibitem{RN578}
J.~Leitão, B.~Meuleman, D.~Van De~Ville, and P.~Vuilleumier, ``Computational imaging during video game playing shows dynamic synchronization of cortical and subcortical networks of emotions,'' \emph{PLOS Biology}, vol.~18, no.~11, p. e3000900, 2020.

\bibitem{RN269}
K.~Scherer, A.~Dieckmann, M.~Unfried, H.~Ellgring, and M.~Mortillaro, ``Investigating appraisal-driven facial expression and inference in emotion communication,'' \emph{Emotion}, 2019.

\bibitem{RN34}
C.~van Reekum, T.~Johnstone, R.~Banse, A.~Etter, T.~Wehrle, and K.~Scherer, ``Psychophysiological responses to appraisal dimensions in a computer game,'' \emph{Cognition and Emotion}, vol.~18, no.~5, pp. 663--688, 2004.

\bibitem{RN54}
S.~Katsigiannis and N.~Ramzan, ``Dreamer: A database for emotion recognition through eeg and ecg signals from wireless low-cost off-the-shelf devices,'' \emph{IEEE Journal of Biomedical and Health Informatics}, vol.~22, no.~1, pp. 98--107, 2018.

\bibitem{RN83}
R.~Subramanian, J.~Wache, M.~K. Abadi, R.~L. Vieriu, S.~Winkler, and N.~Sebe, ``Ascertain: Emotion and personality recognition using commercial sensors,'' \emph{IEEE Transactions on Affective Computing}, vol.~9, no.~2, pp. 147--160, 2018.

\bibitem{RN206}
S.~D. Kreibig, ``Autonomic nervous system activity in emotion: A review,'' \emph{Biological Psychology}, vol.~84, no.~3, pp. 394--421, 2010.

\bibitem{RN477}
P.~Schmidt, A.~Reiss, R.~Dürichen, and K.~V. Laerhoven, ``Wearable-based affect recognition—a review,'' \emph{Sensors}, vol.~19, no.~19, p. 4079, 2019.

\bibitem{RN194}
G.~Chen, X.~Zhang, Y.~Sun, and J.~Zhang, ``Emotion feature analysis and recognition based on reconstructed eeg sources,'' \emph{IEEE Access}, vol.~8, pp. 11\,907--11\,916, 2020.

\bibitem{RN604}
J.~Teo and J.~T. Chia, ``Deep neural classifiers for eeg-based emotion recognition in immersive environments,'' in \emph{2018 International Conference on Smart Computing and Electronic Enterprise (ICSCEE)}, 2018, Conference Proceedings, pp. 1--6.

\bibitem{RN599}
I.~Mavridou, E.~Seiss, M.~Hamedi, E.~Balaguer-Ballester, and C.~Nduka, ``Towards valence detection from emg for virtual reality applications,'' in \emph{12th International Conference on Disability, Virtual Reality and Associated Technologies (ICDVRAT 2018)}.\hskip 1em plus 0.5em minus 0.4em\relax Reading UK: ICDVRAT, University of Reading, September 2018.

\bibitem{RN669}
M.~Perusquía-Hernández, M.~Hirokawa, and K.~Suzuki, ``Spontaneous and posed smile recognition based on spatial and temporal patterns of facial emg,'' in \emph{2017 Seventh International Conference on Affective Computing and Intelligent Interaction (ACII)}.\hskip 1em plus 0.5em minus 0.4em\relax IEEE, 2017, Conference Proceedings, pp. 537--541.

\bibitem{RN664}
J.~T. Cacioppo, R.~E. Petty, M.~E. Losch, and H.~S. Kim, ``Electromyographic activity over facial muscle regions can differentiate the valence and intensity of affective reactions.'' \emph{Journal of personality and social psychology}, vol.~50, no.~2, p. 260, 1986.

\bibitem{RN705}
U.~Dimberg, M.~Thunberg, and K.~Elmehed, ``Unconscious facial reactions to emotional facial expressions,'' vol.~11, no.~1, pp. 86--89, 2000.

\bibitem{RN627}
W.~Sato, K.~Murata, Y.~Uraoka, K.~Shibata, S.~Yoshikawa, and M.~Furuta, ``Emotional valence sensing using a wearable facial emg device,'' \emph{Scientific Reports}, vol.~11, no.~1, p. 5757, 2021.

\bibitem{RN494}
G.~Bernal, T.~Yang, A.~Jain, and P.~Maes, ``Physiohmd: A conformable, modular toolkit for collecting physiological data from head-mounted displays,'' in \emph{Proceedings of the 2018 ACM International Symposium on Wearable Computers}, 2018, p. 160–167.

\bibitem{RN666}
V.~Kehri and R.~Awale, ``A facial emg data analysis for emotion classification based on spectral kurtogram and cnn,'' \emph{International Journal of Digital Signals and Systems, Smart}, vol.~4, no. 1-3, pp. 50--63, 2020.

\bibitem{RN610}
J.~Perdiz, G.~Pires, and U.~J. Nunes, ``Emotional state detection based on emg and eog biosignals: A short survey,'' in \emph{2017 IEEE 5th Portuguese Meeting on Bioengineering (ENBENG)}, Conference Proceedings, pp. 1--4.

\bibitem{RN668}
L.~Boot, ``Facial expressions in eeg/emg recordings,'' Thesis, 2009.

\bibitem{RN678}
L.~Schilbach, S.~B. Eickhoff, A.~Mojzisch, and K.~Vogeley, ``What's in a smile? neural correlates of facial embodiment during social interaction,'' \emph{Social Neuroscience}, vol.~3, no.~1, pp. 37--50, 2008.

\bibitem{RN703}
T.~Luong, A.~Lecuyer, N.~Martin, and F.~Argelaguet, ``A survey on affective and cognitive vr,'' \emph{IEEE Transactions on Visualization and Graphics, Computer}, 2021.

\bibitem{RN30}
C.~Bassano, G.~Ballestin, E.~Ceccaldi, F.~I. Larradet, M.~Mancini, E.~Volta, and R.~Niewiadomski, ``A vr game-based system for multimodal emotion data collection.''\hskip 1em plus 0.5em minus 0.4em\relax New York, NY, USA: Association for Computing Machinery, 2019.

\bibitem{RN605}
I.~Shumailov and H.~Gunes, ``Computational analysis of valence and arousal in virtual reality gaming using lower arm electromyograms,'' in \emph{2017 Seventh International Conference on Affective Computing and Intelligent Interaction (ACII)}, 2017, Conference Proceedings, pp. 164--169.

\bibitem{RN190}
V.~Shuman, K.~Schlegel, and K.~Scherer, \emph{Geneva Emotion Wheel Rating Study}, 2015.

\bibitem{RN706}
W.~Sato, T.~Kochiyama, and S.~Yoshikawa, ``Physiological correlates of subjective emotional valence and arousal dynamics while viewing films,'' \emph{Biological Psychology}, vol. 157, p. 107974, 2020.

\bibitem{RN660}
M.~Gnacek, J.~Broulidakis, I.~Mavridou, M.~Fatoorechi, E.~Seiss, T.~Kostoulas, E.~Balaguer-Ballester, I.~Kiprijanovska, C.~Rosten, and C.~Nduka, ``emteqpro—fully integrated biometric sensing array for non-invasive biomedical research in virtual reality,'' vol.~3, 2022.

\bibitem{RN651}
D.~S. Elvitigala, D.~J.~C. Matthies, and S.~Nanayakkara, ``Stressfoot: Uncovering the potential of the foot for acute stress sensing in sitting posture,'' vol.~20, no.~10, p. 2882, 2020.

\bibitem{RN302}
S.~Lamb and K.~Kwok, ``Mssq-short norms may underestimate highly susceptible individuals,'' \emph{Human Factors: The Journal of the Human Factors and Ergonomics Society}, vol.~57, 2014.

\bibitem{RN702}
J.~W. Peirce, ``Psychopy—psychophysics software in python,'' \emph{Journal of neuroscience methods}, vol. 162, no. 1-2, pp. 8--13, 2007.

\bibitem{RN699}
T.~Mullen, \emph{Mastering blender}.\hskip 1em plus 0.5em minus 0.4em\relax John Wiley \& Sons, 2011.

\bibitem{RN698}
\BIBentryALTinterwordspacing
 [Online]. Available: \url{https://www.harfang3d.com/en_US/}
\BIBentrySTDinterwordspacing

\bibitem{RN248}
C.~E. Izard, ``Basic emotions, natural kinds, emotion schemas, and a new paradigm,'' vol.~2, no.~3, pp. 260--280, 2007.

\bibitem{RN249}
------, ``Emotion theory and research: Highlights, unanswered questions, and emerging issues,'' \emph{Annual review of psychology}, vol.~60, pp. 1--25, 2009.

\bibitem{RN25}
G.~Mohammadi, K.~Lin, and P.~Vuilleumier, ``Towards understanding emotional experience in a componential framework,'' in \emph{2019 8th International Conference on Affective Computing and Intelligent Interaction (ACII)}, 2019, Conference Proceedings, pp. 123--129.

\bibitem{RN704}
J.~A. Russell, M.~Lewicka, and T.~Niit, ``A cross-cultural study of a circumplex model of affect,'' \emph{Journal of personality and psychology, social}, vol.~57, no.~5, p. 848, 1989.

\bibitem{RN230}
X.~S. P.~P. Alexandre~Schaefer, Frédéric~Nils, ``Assessing the effectiveness of a large database of emotion-eliciting films: A new tool for emotion researchers.'' \emph{Cognition \& Emotion}, vol.~24, pp. 1153--1172, 2010.

\bibitem{RN581}
J.~R. Fontaine, K.~R. Scherer, E.~B. Roesch, and P.~C. Ellsworth, ``The world of emotions is not two-dimensional,'' \emph{Psychol Sci}, vol.~18, no.~12, pp. 1050--7, 2007.

\bibitem{RN189}
K.~R. Scherer, V.~Shuman, J.~J.~R. Fontaine, and C.~Soriano, \emph{The GRID meets the Wheel: Assessing emotional feeling via self-report1}.\hskip 1em plus 0.5em minus 0.4em\relax Oxford: Oxford University Press, 2013.

\bibitem{RN572}
G.~Mohammadi, D.~Van De~Ville, and P.~Vuilleumier, ``Brain networks subserving functional core processes of emotions identified with componential modeling,'' \emph{Cerebral Cortex}, vol.~33, no.~12, pp. 7993--8010, 2023.

\bibitem{RN582}
U.~Beermann, G.~Hosoya, I.~Schindler, K.~R. Scherer, M.~Eid, V.~Wagner, and W.~Menninghaus, ``Dimensions and clusters of aesthetic emotions: A semantic profile analysis,'' vol.~12, no. 1949, 2021.

\bibitem{RN708}
R.~Reisenzein and C.~Spielhofer, ``Subjectively salient dimensions of emotional appraisal,'' \emph{Motivation and Emotion}, vol.~18, no.~1, pp. 31--77, 1994.

\bibitem{RN709}
J.~A. Russell, \emph{Reading emotions from and into faces: Resurrecting a dimensional-contextual perspective}, ser. Studies in emotion and social interaction, 2nd series.\hskip 1em plus 0.5em minus 0.4em\relax Paris, France: Editions de la Maison des Sciences de l'Homme, 1997, pp. 295--320.

\bibitem{RN696}
M.~Gjoreski, I.~Kiprijanovska, S.~Stankoski, I.~Mavridou, M.~J. Broulidakis, H.~Gjoreski, and C.~Nduka, ``Facial emg sensing for monitoring affect using a wearable device,'' \emph{Scientific reports}, vol.~12, no.~1, pp. 1--12, 2022.

\bibitem{RN667}
P.~Ekman and W.~V. Friesen, ``Facial action coding system,'' \emph{Environmental Psychology \& Nonverbal Behavior}, 1978.

\bibitem{RN207}
J.~Kory and S.~K. D'Mello, ``Affect elicitation for affective computing,'' Conference Proceedings.

\bibitem{RN7}
M.~Granato, D.~Gadia, D.~Maggiorini, and L.~A. Ripamonti, ``An empirical study of players’ emotions in vr racing games based on a dataset of physiological data,'' \emph{Multimedia Tools and Applications}, 2020.

\bibitem{RN694}
M.~J. Kim, A.~M. Mattek, R.~H. Bennett, K.~M. Solomon, J.~Shin, and P.~J. Whalen, ``Human amygdala tracks a feature-based valence signal embedded within the facial expression of surprise,'' \emph{Journal of Neuroscience}, vol.~37, no.~39, pp. 9510--9518, 2017.

\bibitem{RN662}
M.~G. Frank, P.~Ekman, and W.~V. Friesen, ``Behavioral markers and recognizability of the smile of enjoyment,'' \emph{Journal of personality and psychology, social}, vol. 64 1, pp. 83--93, 1993.

\bibitem{RN707}
C.~F. Lima, P.~Arriaga, A.~Anikin, A.~R. Pires, S.~Frade, L.~Neves, and S.~K. Scott, ``Authentic and posed emotional vocalizations trigger distinct facial responses,'' \emph{Cortex}, vol. 141, pp. 280--292, 2021.

\bibitem{RN613}
A.~Gruebler, V.~Berenz, and K.~Suzuki, ``Emotionally assisted human–robot interaction using a wearable device for reading facial expressions,'' \emph{Advanced Robotics}, vol.~26, no.~10, pp. 1143--1159, 2012.

\bibitem{RN665}
L.~Inzelberg, D.~Rand, S.~Steinberg, M.~David-Pur, and Y.~Hanein, ``A wearable high-resolution facial electromyography for long term recordings in freely behaving humans,'' \emph{Scientific reports}, vol.~8, no.~1, pp. 1--9, 2018.

\bibitem{RN657}
V.~Chandra, A.~Priyarup, and D.~Sethia, ``Comparative study of physiological signals from empatica e4 wristband for stress classification,'' in \emph{International Conference on Advances in Computing and Data Sciences}.\hskip 1em plus 0.5em minus 0.4em\relax Springer, Conference Proceedings, pp. 218--229.

\bibitem{RN659}
A.~Savitzky and M.~J.~E. Golay, ``Smoothing and differentiation of data by simplified least squares procedures,'' \emph{Analytical Chemistry}, vol.~36, no.~8, pp. 1627--1639, 1964.

\bibitem{RN700}
M.~Sevil, M.~Rashid, I.~Hajizadeh, M.~R. Askari, N.~Hobbs, R.~Brandt, M.~Park, L.~Quinn, and A.~Cinar, ``Discrimination of simultaneous psychological and physical stressors using wristband biosignals,'' \emph{Computer Methods and Biomedicine, Programs in}, vol. 199, p. 105898, 2021.

\bibitem{RN503}
N.~V. Chawla, K.~W. Bowyer, L.~O. Hall, and W.~P. Kegelmeyer, ``Smote: synthetic minority over-sampling technique,'' \emph{Journal of artificial intelligence research}, vol.~16, pp. 321--357, 2002.

\bibitem{RN111}
J.~Yih, L.~D. Kirby, and C.~A. Smith, ``Profiles of appraisal, motivation, and coping for positive emotions,'' \emph{Cognition and Emotion}, vol.~34, no.~3, pp. 481--497, 2020.

\end{thebibliography}

\vspace{-1cm}
\begin{IEEEbiography}[{\includegraphics[width=1in,height=1.25in,clip,keepaspectratio]{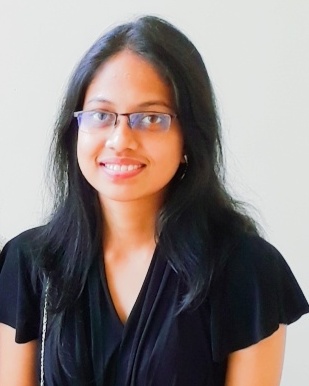}}]{Rukshani Somarathna}
is with the UNSW, Sydney, where she is a Doctoral student at the School of Computer Science and Engineering. She received her BSc Hons in Information Technology from the University of Moratuwa, Sri Lanka, in 2017. From 2017 to 2019, she was a Software Engineer who was involved in the development of a Healthcare System. Her current research interests include Affective Computing, Artificial Intelligence, Signal Processing, and Virtual Reality.
\end{IEEEbiography}

\begin{IEEEbiography}[{\includegraphics[width=1in,height=1.25in, clip,keepaspectratio]{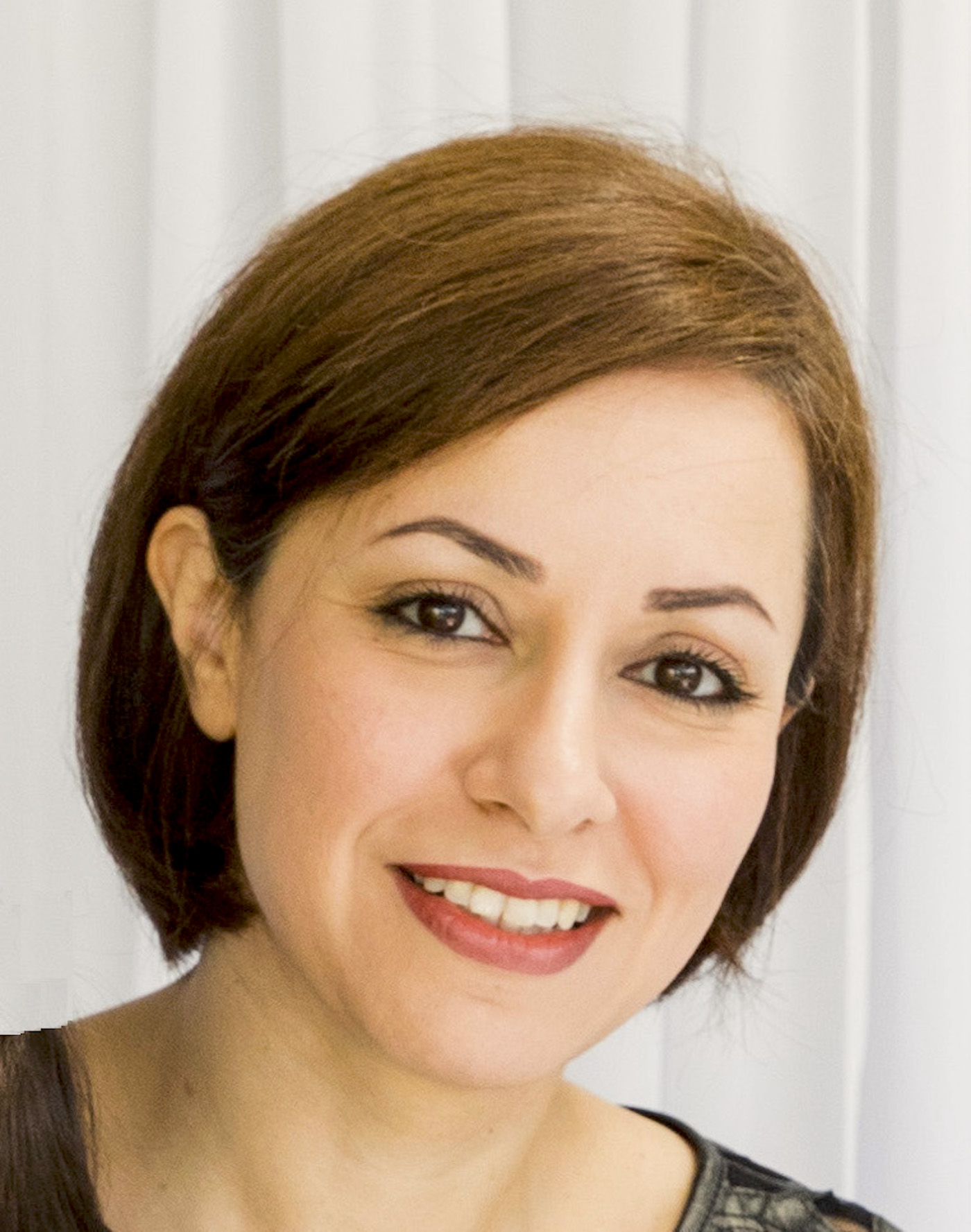}}]{Gelareh Mohammadi} is a Senior Lecturer (Assistant Prof.) and head of Human-Centred Computing (HCC) research group at the School of Computer Science and Engineering, UNSW, Sydney. Her main research interests are in Affective Computing and  Behavior Analysis with the aim of better understanding humans and designing human-centred technologies. She received her Ph.D. in Electrical Engineering from EPFL, Switzerland. She was a post-doctoral researcher at Idiap Research Institute and later at Laboratory for Neurology and Imaging of Cognition, University of Geneva. She has investigated approaches for personality perception, emotion recognition from a componential perspective, and understanding the neural basis of emotions. Gelareh has authored and co-authored more than 70 publications, including several book chapters. She was a recipient of the ”Google Anita Borg award” (a.k.a. Women Techmakers) in 2013 and has been selected as the field leader in the human-computer-interaction domain in The Australian’s Research Report, 2019. 
\end{IEEEbiography}

\end{document}


\title{Exploring Emotions in Multi-componential Space using Interactive VR Games}

\author{Rukshani~Somarathna,~\IEEEmembership{Member,~IEEE,}
        and~Gelareh~Mohammadi,~\IEEEmembership{Member,~IEEE}
}

\maketitle

\IEEEraisesectionheading{\section{Supplementary Materials}\label{sec:introduction}}

\begin{table}[htbp]
\caption{Commercially available twenty-seven VR games from the Steam platform used in the study with the pre-tagged dominant emotion.}
\label{tab:PreTaggedVRGames}
\centering
{
    \begin{tabular}{p{15mm}p{55mm}}
    \textbf{Dominant emotion} & 
    \textbf{Games} \\ \hline
    Fear & The Brookhaven Experiment, Kitty Rescue, Propagation VR, Richies Plank Experience \\ 
    Interest & Doctor Who: the runway, Toran, Waltz of the Wizard (Legacy) \\ 
    Amusement & Angry Birds VR: Isle of Pigs, Fruit Ninja VR, Moss, Fujii \\ 
    Joy & Cartoon Network Journeys VR, Beat Saber, Futurejam, Tilt Brush, MSI Electric City: Core Assault \\ 
    Contentment & A Fisherman's Tale, Virus Popper, Egg Time \\ 
    Admiration & Ocean Rift \\ 
    Relief & Meditation VR, Catch \& Release, Google Earth VR, Cosmic Flow \\
    Compassion & The Book of Distance \\
    Sadness & Everybody's sad \\
    Angry & Deadly Hunter VR \\\hline
    \end{tabular}
    }
\end{table}

\begin{table}[htbp]
  \caption{Geneva Emotion Wheel questionnaire}
  \label{tab:GEWSurveyTable}
  \centering
  \begin{tabular}{p{40mm}p{40mm}}
    \multicolumn{2}{c}{\textbf{While playing this game, I felt…}} \\ \hline
    Interest/Alert/Curious & Sadness/Downhearted/Unhappy \\
    Amusement/Awe/Wonder & Guilt/Repentant \\
    Pride/Confident/Self-assured & Regret \\
    Joy/Glad/Happy & Shame/Humiliated/Disgraced \\
    Pleasure & Disappointment \\
    Contentment & Fear/Scared/Afraid \\
    Admiration &  Disgust/Distaste \\
    Love/Closeness/Trust & Contempt \\
    Relief & Hate/Distrust \\
    Compassion & Angry/Irritated/Annoyed \\ \hline
\end{tabular}
\end{table}

\begin{table}[htbp]
  \caption{CoreGRID questionnaire}
  \label{tab:CoreGRIDSurvey}
  \centering
  \begin{tabular}{p{60mm}p{10mm}}
    \multicolumn{2}{c}{\textbf{While playing this game, did you/r…}} \\
    \textbf{CoreGRID item} &	
    \textbf{Component} \\ \hline
    experience an intense emotional state? & 	Feeling \\ 
    experience a prolonged heightened emotional state? &	Feeling \\
    feel good? &	Feeling \\
    feel tired? &	Feeling \\
    feel energised? &	Feeling \\
    feel calm? &	Feeling \\
    feel bad? &	Feeling \\
    feel weak? &	Feeling \\ \hline
    feel nervous or weak in the knees? &	Physiology \\ 
    stomach clench up? &	Physiology \\
    experience an elevated heart rate? &	Physiology \\
    experience muscle tension? &	Physiology \\
    feel your breathing slowing down? &	Physiology \\
    feel your breathing getting faster? &	Physiology \\
    feel warm? &	Physiology \\
    sweat? &	Physiology \\ \hline
    smile? &	Expression \\
    jaw drop? &	Expression \\
    eyebrows go up? &	Expression \\
    frown? &	Expression \\
    shut your eyes? &	Expression \\
    cry? &	Expression \\
    shout or exclaim? &	Expression \\
    experience speech disturbances? &	Expression \\ \hline
    want the experience to continue? &	Motivation \\
    want the experience to stop? &	Motivation \\
    want to reverse what was happening? &	Motivation \\
    feel motivated to overcome an obstacle? &	Motivation \\
    feel bored or unmotivated? &	Motivation \\
    feel an urge to lash out verbally or physically? &	Motivation \\
    feel an urge to resist or oppose someone or something? &	Motivation \\
    feel motivated to tackle the situation at hand? &	Motivation \\
    feel that you want to run away (virtually)? &	Motivation \\
    want to sing and dance? &	Motivation \\ \hline
    feel that the experience was predictable? &	Appraisal \\
    think that the experience caused negative or undesirable consequences for you? & Appraisal \\
    feel the experience caused by chance? &	Appraisal \\
    feel that the experience violated laws/social norms? &	Appraisal \\
    feel the events occurred suddenly/without warning? &	Appraisal \\
    feel the experience demanded an immediate response? &	Appraisal \\
    event was caused by a virtual agent? &	Appraisal \\
    feel the event was important for and relevant to your goals or needs? &	Appraisal \\
    feel like the experience was uncontrollable? &	Appraisal \\
    think the experience was pleasant? &	Appraisal \\
    feel that the experience was unpredictable? &	Appraisal \\
    feel that there was no urgency in the situation? &	Appraisal \\
    feel in control over the outcome? &	Appraisal \\
    feel that the outcomes were a result of your behaviour? &	Appraisal \\
    feel that the experience was incongruent with your standards/ideals? &	Appraisal \\
    feel that you had the power over the consequences of the event? &	Appraisal \\
    feel powerless in the situation? &	Appraisal \\
  \hline
\end{tabular}
\end{table}

\pgfplotstableread{
    Label           1-	    2-	    3-	    4-	    5-
    Interest	    66.0	123.0	227.0	351.0	274.0
    Amusement	    142.0	160.0	261.0	292.0	186.0
    Pride	        315.0	190.0	285.0	171.0	80.0
    Joy	            152.0	159.0	270.0	264.0	196.0
    Pleasure	    154.0	155.0	274.0	286.0	172.0
    Contentment	    254.0	162.0	258.0	241.0	126.0
    Admiration	    436.0	155.0	232.0	147.0	71.0
    Love	        501.0	189.0	167.0	94.0	90.0
    Relief	        460.0	180.0	205.0	144.0	52.0
    Compassion	    601.0	145.0	152.0	110.0	33.0
    Sadness	        813.0	123.0	66.0	31.0	8.0
    Guilt	        892.0	90.0	39.0	18.0	2.0
    Regret	        775.0	137.0	82.0	40.0	7.0
    Shame	        848.0	118.0	54.0	19.0	2.0
    Disappointment	656.0	183.0	99.0	83.0	20.0
    Fear	        757.0	123.0	67.0	47.0	47.0
    Disgust	        851.0	93.0	52.0	27.0	18.0
    Contempt	    754.0	129.0	89.0	52.0	17.0
    Hate	        837.0	113.0	49.0	28.0	14.0
    Angry	        739.0	141.0	90.0	58.0	13.0
}\EmoCompVRGEWLikertdatadistribution

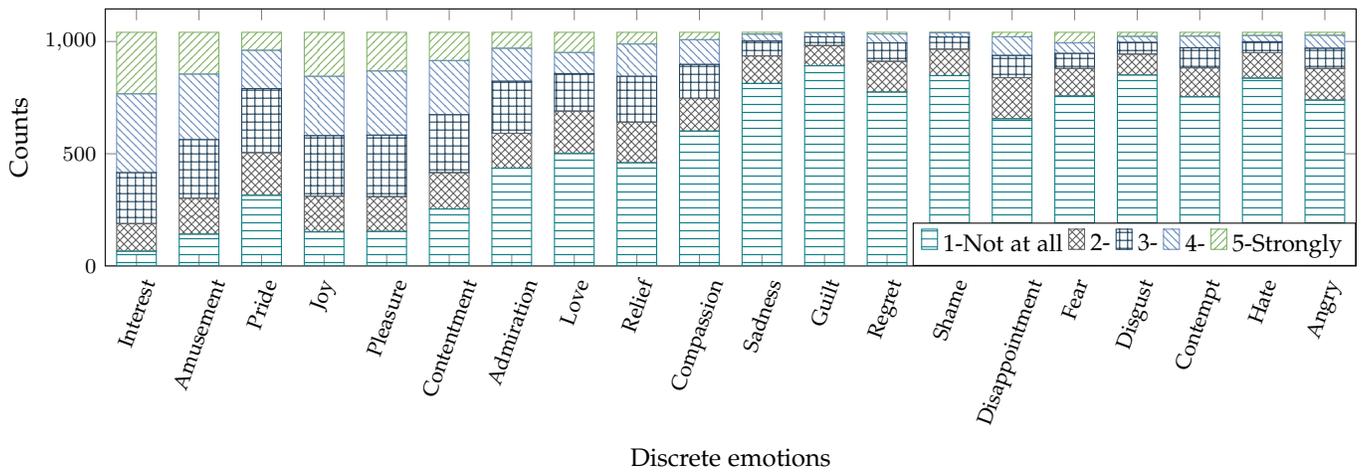
\begin{figure*}[t]
    \begin{tikzpicture}
        \begin{axis}[
            ybar stacked,
            ymin=0,
            bar width=15pt,
            height=5cm,
            width=\linewidth,
            xtick=data,
            xtick distance=1,
            xlabel=Discrete emotions,
            yticklabel style={align=center,font=\footnotesize},
            xticklabel style={align=center,font=\footnotesize, rotate=70},
            ylabel=Counts,
            enlarge x limits={abs=0.5},
            legend cell align={left},
            legend columns = 5,
            legend style={at={(1,0)},anchor=south east, font=\small},
            xticklabels from table={\EmoCompVRGEWLikertdatadistribution}{Label},
        ]
        	 					
        \addplot [AppraisalColor, pattern=horizontal lines, pattern color=AppraisalColor] table [y=1-, meta=Label, x expr=\coordindex] {\EmoCompVRGEWLikertdatadistribution};
        \addlegendentry{1-Not at all}
        \addplot [MotivationColor, pattern=crosshatch, pattern color=MotivationColor] table [y=2-, meta=Label, x expr=\coordindex] {\EmoCompVRGEWLikertdatadistribution};
        \addlegendentry{2-}
        \addplot [PhysiologyColor, pattern=grid, pattern color=PhysiologyColor] table [y=3-, meta=Label, x expr=\coordindex] {\EmoCompVRGEWLikertdatadistribution};
        \addlegendentry{3-}
        \addplot [ExpressionColor, pattern=north west lines, pattern color=ExpressionColor]  table [y=4-, meta=Label, x expr=\coordindex] {\EmoCompVRGEWLikertdatadistribution};
        \addlegendentry{4-}
        \addplot [FeelingColor, pattern=north east lines, pattern color=FeelingColor] table [y=5-, meta=Label, x expr=\coordindex] {\EmoCompVRGEWLikertdatadistribution};
        \addlegendentry{5-Strongly}
        
        \end{axis}
    \end{tikzpicture}
    \caption{Distribution of scorings in 20 emotions. Colour scales correspond to the proportion of samples.}
    \label{fig:EmoCompVR_DistributionGEW}
\end{figure*}

\begin{figure*}[htbp]
    \centering
    \includegraphics[trim=0.5cm 0.5cm 5cm 8cm, clip, width=0.8\linewidth]{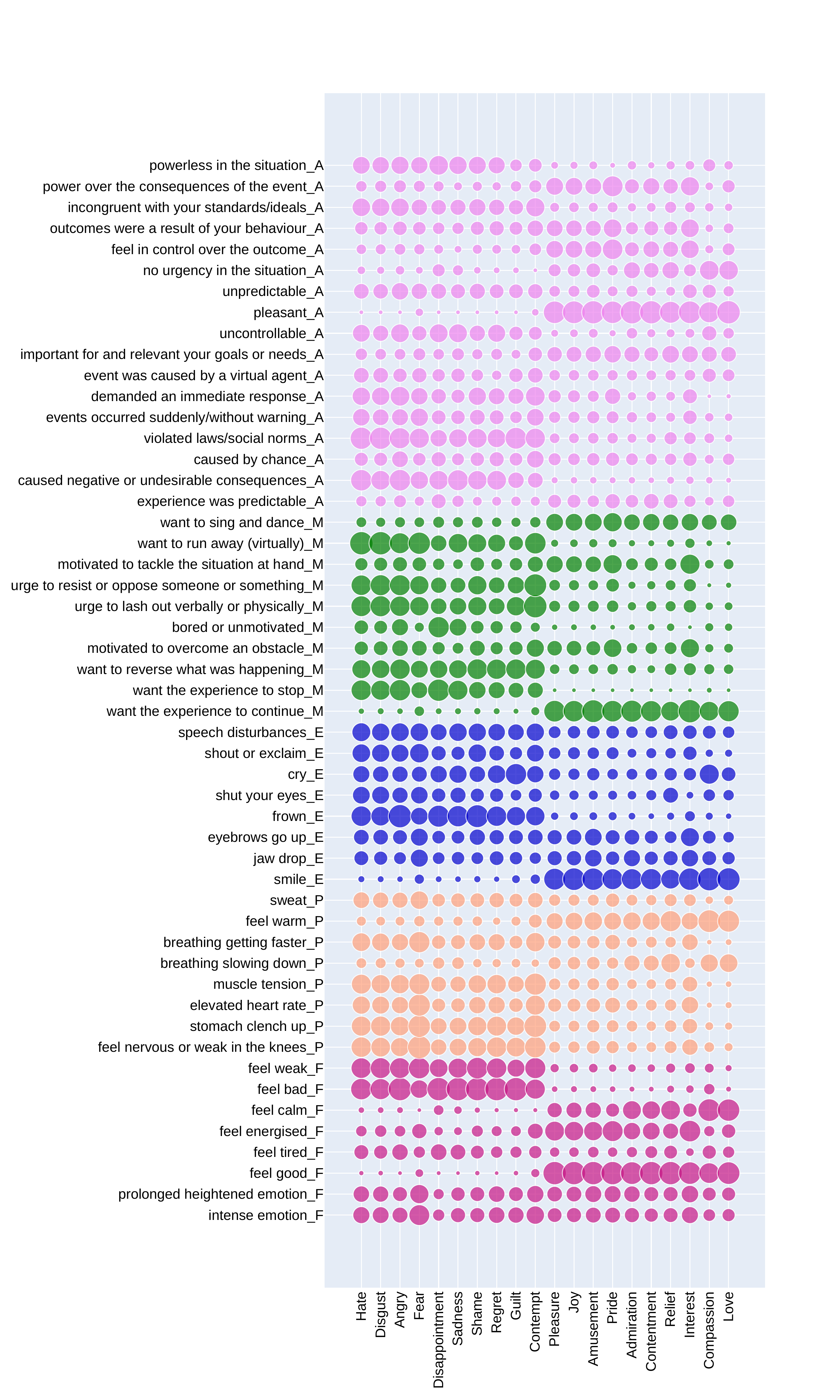}
    \caption{Emotion profiles in CoreGRID space derived from hierarchical clustering. The size of the bubble is proportional to the value of the weighted average CPM profile for each emotion term after scaling for visualisation.}
    \label{fig:FigureS1_HC_GRIDProfiles}
\end{figure*}

\begin{figure}[htbp]
    \centering
    \includegraphics[trim=0.5cm 0.5cm 5cm 6cm, clip, width=1\linewidth]{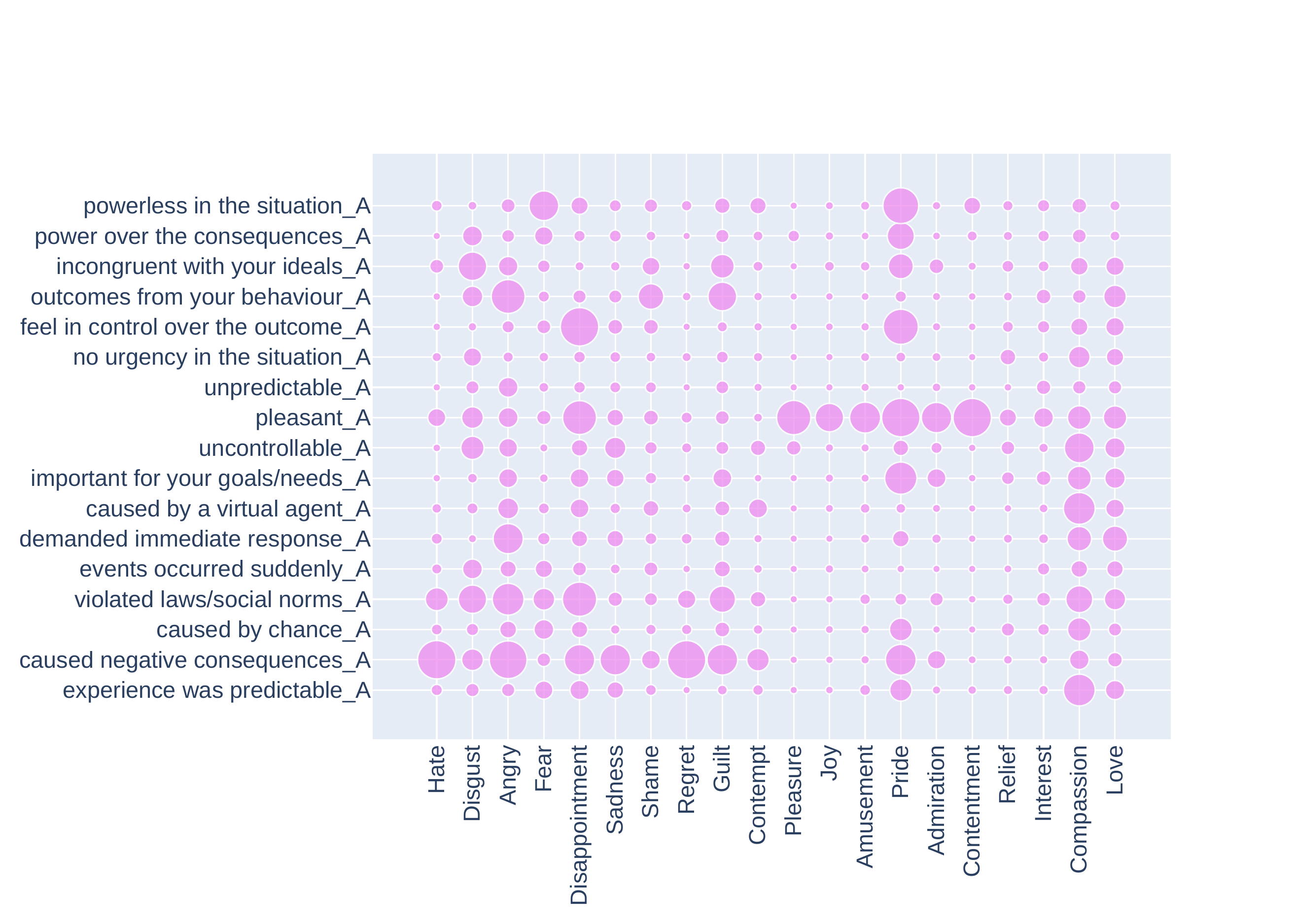}
    \caption{Appraisal component feature importance obtained from the LGBM model in section 4.5.2. The size of the bubble is proportional to the feature's importance. Results have been scaled for visualisation.}
    \label{fig:FigureS2_FI_Appraisal}
\end{figure}

\begin{figure}[htbp]
    \centering
    \includegraphics[trim=0.5cm 0.5cm 5cm 6cm, clip, width=1\linewidth]{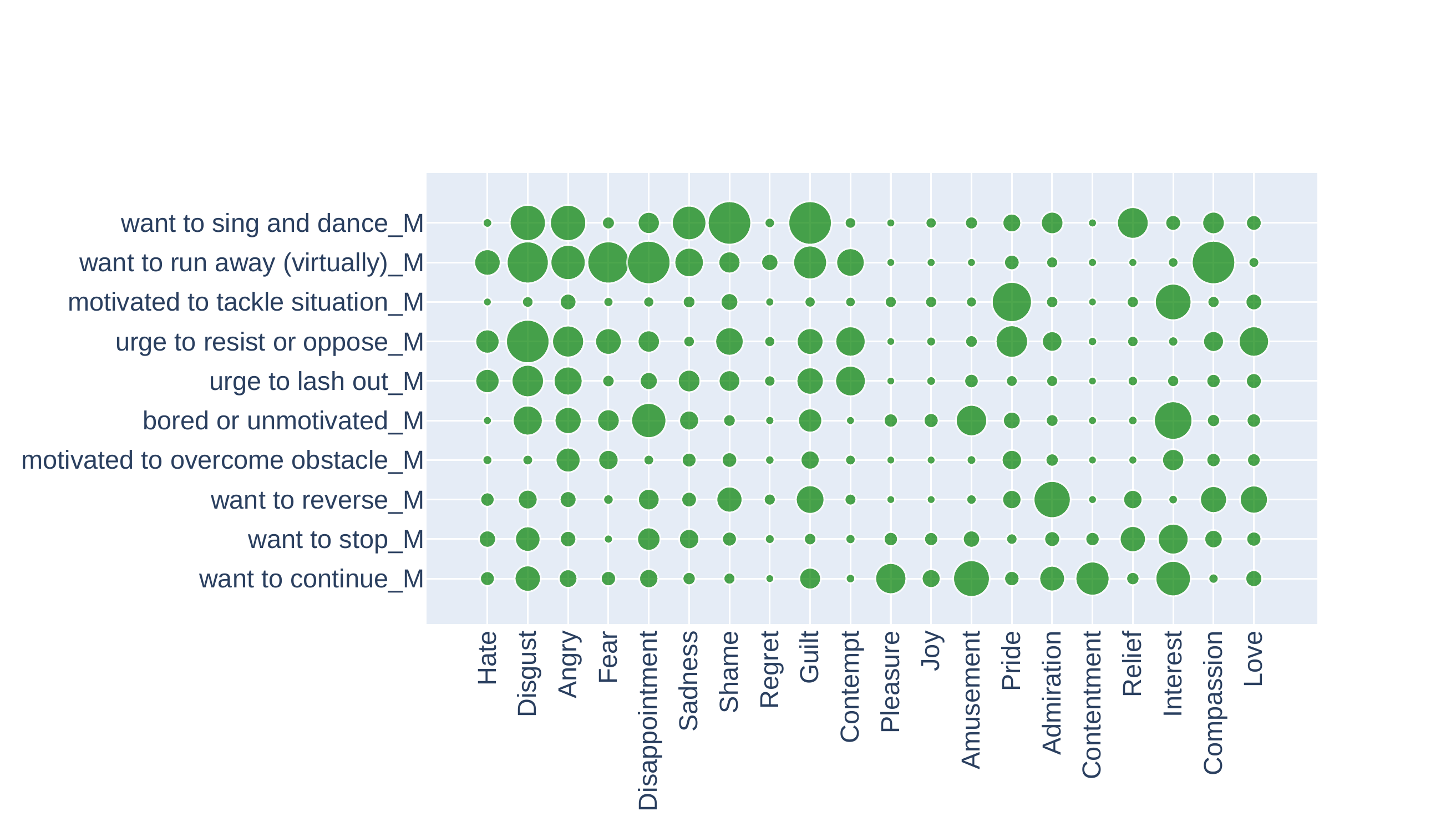}
    \caption{Motivation component feature importance obtained from the LGBM model in section 4.5.2. The size of the bubble is proportional to the feature's importance. Results have been scaled for visualisation.}
    \label{fig:FigureS3_FI_Motivation}
\end{figure}

\begin{figure}[htbp]
    \centering
    \includegraphics[trim=0.5cm 0.5cm 5cm 4cm, clip, width=1\linewidth]{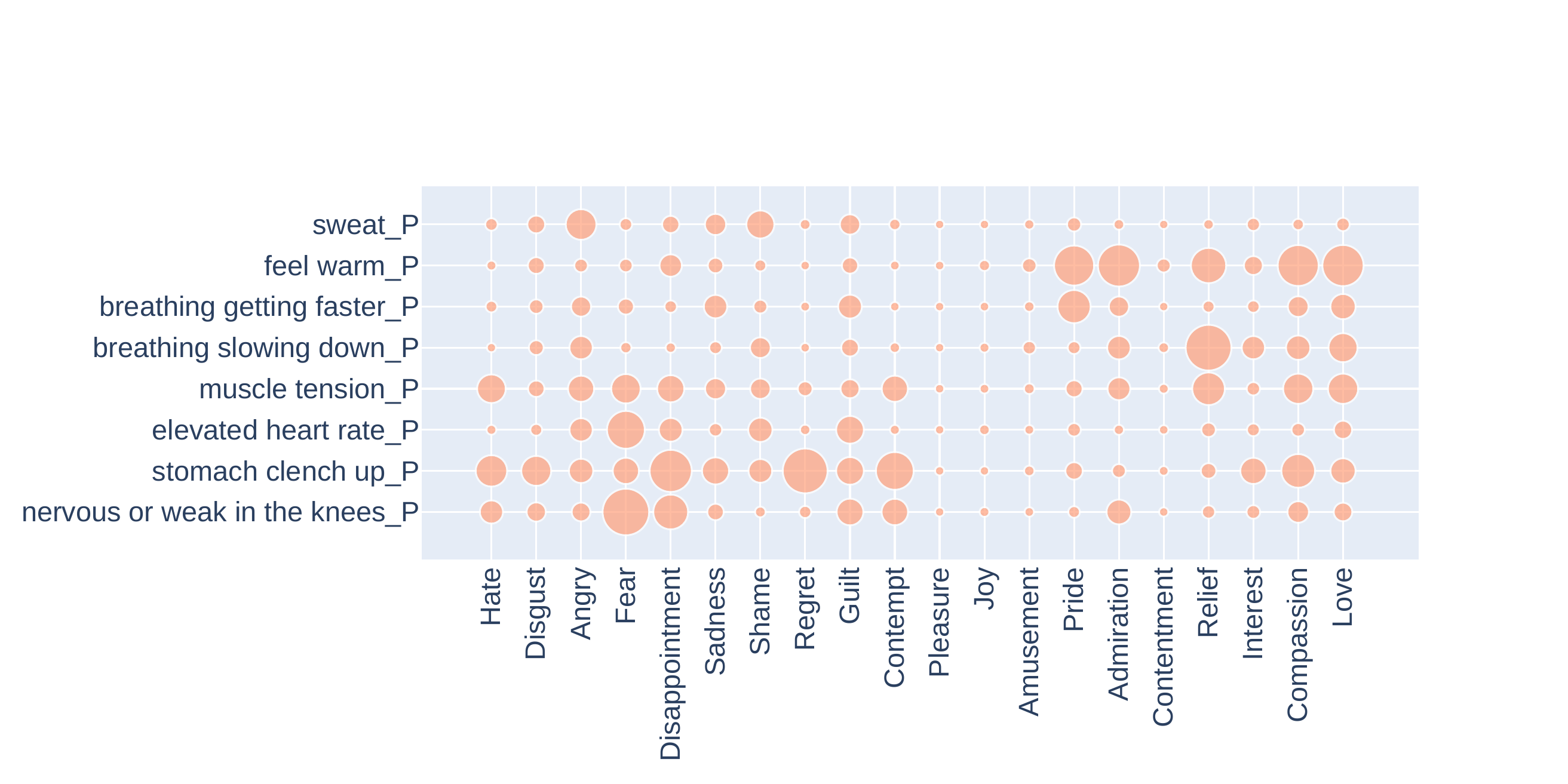}
    \caption{Physiology component feature importance obtained from the LGBM model in section 4.5.2. The size of the bubble is proportional to the feature's importance. Results have been scaled for visualisation.}
    \label{fig:FigureS4_FI_Physiology}
\end{figure}

\begin{figure}[!t]
    \centering
    \includegraphics[trim=0.5cm 0.5cm 5cm 5cm, clip, width=1\linewidth]{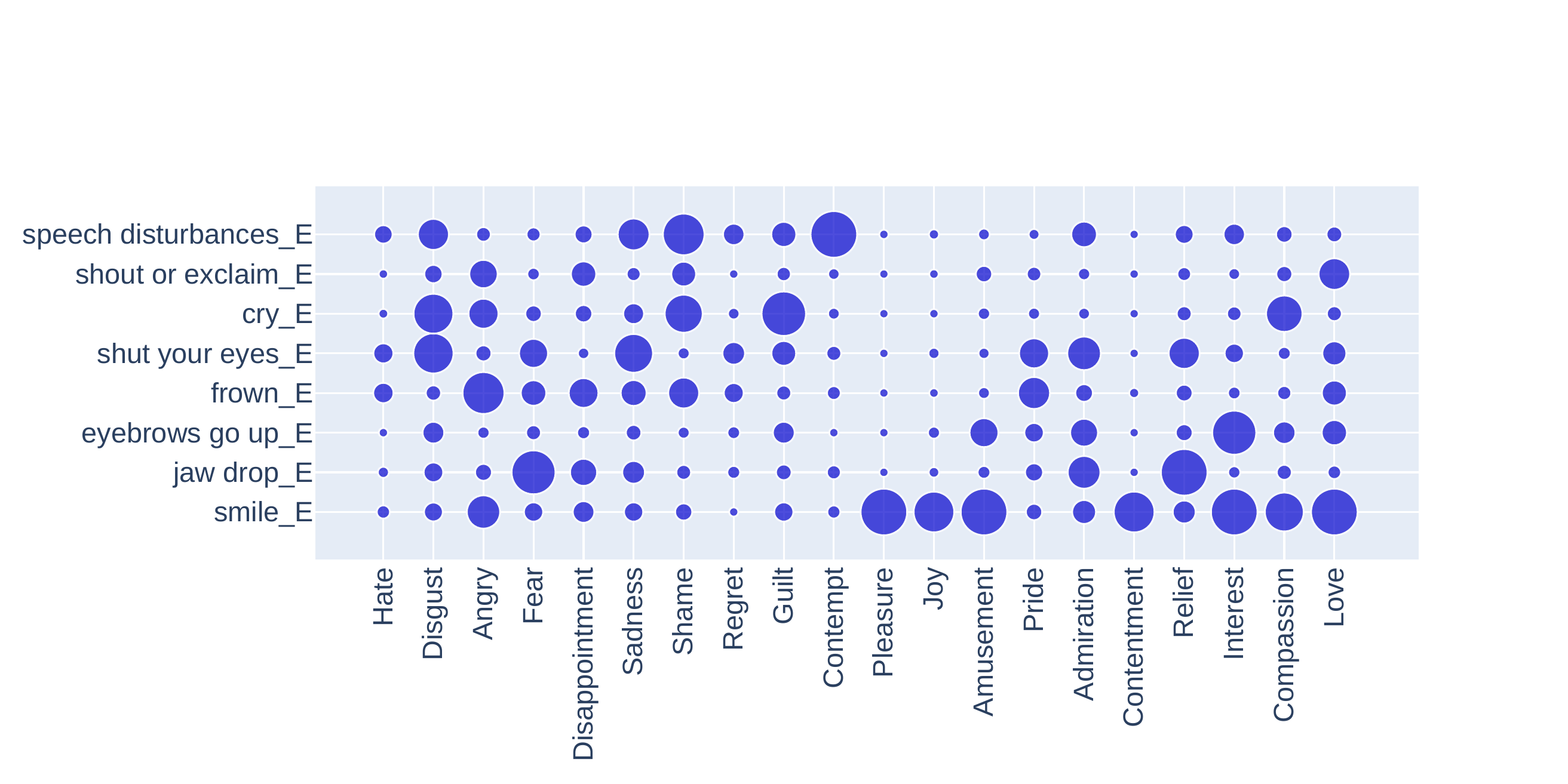}
    \caption{Expression component feature importance obtained from the LGBM model in section 4.5.2. The size of the bubble is proportional to the feature's importance. Results have been scaled for visualisation.}
    \label{fig:FigureS5_FI_Expression}
\end{figure}

\vspace{-100cm}

\begin{figure}[htbp]
    \centering
    \includegraphics[trim=0.5cm 0.5cm 5cm 4cm, clip, width=1\linewidth]{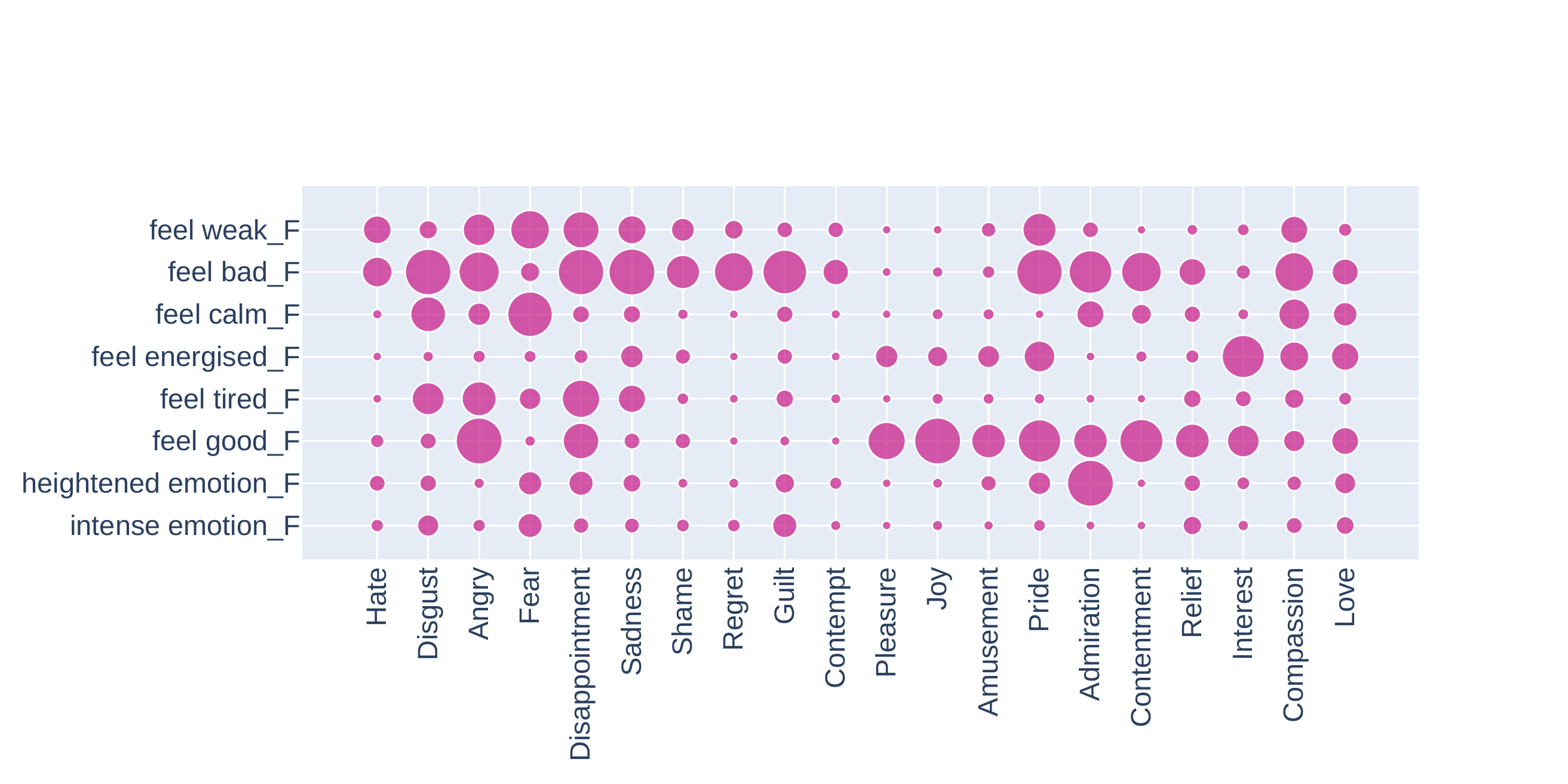}
    \caption{Feeling component feature importance obtained from the LGBM model in section 4.5.2. The size of the bubble is proportional to the feature's importance. Results have been scaled for visualisation.}
    \label{fig:FigureS6_FI_Feeling}
\end{figure}